\documentclass[aps,prd,amsmath,amssymb,superscriptaddress,showkeys,showpacs,twocolumn,floatfix]{revtex4-1}

\usepackage{amsmath,amssymb,mathtools}	
\usepackage{graphicx,color,subfigure}
\usepackage{float}
\usepackage{appendix}
\usepackage{hyperref}
\usepackage{booktabs}

\begin{document}
\title{Gravitational wave and dark matter from Axion-Higgs string}

\author{Yongtao Jia}
\affiliation{~Department of Physics and Chongqing Key Laboratory for Strongly Coupled Physics, Chongqing University, Chongqing 401331, P. R. China}

\author{Ligong Bian\thanks{Corresponding Author.}}
\email{lgbycl@cqu.edu.cn}
\affiliation{~Department of Physics and Chongqing Key Laboratory for Strongly Coupled Physics, Chongqing University, Chongqing 401331, P. R. China}
\affiliation{~ Center for High Energy Physics, Peking University, Beijing 100871, China}

\begin{abstract}
    Axions have long been considered plausible candidates for dark matter. The axion dark matter emitted from cosmic strings after the Peccei-Quinn (PQ) symmetry breaking in the early Universe was extensively simulated. In this work, we study dark matter and gravitational waves through the lattice simulation of the Axion-Higgs string. We gave the dark matter overproduction and the Big Bang nucleosynthesis bounds on the axion decay constant $f_a$ and the axion mass $m_a$ for axion-like particles, and found that the predicted gravitational wave spectra cannot be probed by the dataset of the current pulsar timing array experiments. 
\end{abstract}

\maketitle
\section{introduction} 
The QCD axion was originally proposed to dynamically solve the strong CP problem \cite{Peccei:1977hh,Peccei:1977ur,Weinberg:1977ma,Wilczek:1977pj,Kim:1979if,Shifman:1979if,Zhitnitsky:1980tq,Dine:1981rt}, which was also considered as one of the most motivated dark matter (DM) candidate~\cite{Preskill:1982cy,Abbott:1982af,Dine:1982ah}. The axion arises as a pseudo-Nambu-Goldstone boson after the Peccei-Quinn(PQ) symmetry breaking at high energy scale ($v_a$) connected the axion decay constant ($f_a$) through $v_a=f_a/N_{DW}$ with the $N_{DW}$ being the domain wall number after the QCD phase transition~\cite{DiLuzio:2020wdo}. Supposing the global PQ symmetry is broken after reheating following the cosmic inflation, known as the post-inflationary scenario, one has random initial axion field distribution in uncorrelated horizon-size regions that leads to the formation of axion strings~\cite{Kibble:1976sj,Vilenkin:1982ks}. 

Previous studies show that global strings mostly decay to particles~\cite{Saurabh:2020pqe,Baeza-Ballesteros:2023say}. 
Axion strings after formation keep the number density per Hubble patch with the scaling parameter being near constant and release their energy by radiating free axions and gravitational waves until QCD scale~\cite{Yamaguchi_1999,Hiramatsu_2011},  and the ratio between the gravitational waves (GWs) radiation and particles was given in Ref.~\cite{Baeza-Ballesteros:2023say}.
Recent lattice simulations gave the log dependence of the scaling parameter on the Hubble parameter~\cite{Gorghetto_2018,Gorghetto:2020qws,Buschmann_2022,Benabou:2023ghl,Saikawa:2024bta,Kim:2024wku}. The scaling property of such a string system is crucial to determine the DM relic abundance possibility and the GWs ~\cite{Gorghetto:2021fsn}. The current pulsar timing array (PTA)
collaborations~\cite{EPTA:2023fyk,Reardon:2023gzh,NANOGrav:2023gor,Xu:2023wog} have announced possible evidence for the stochastic GW background, that may arise from cosmic strings (local string)~\cite{Kitajima:2023vre,Bian:2023dnv,Basilakos:2023xof,Wang:2023len,Ellis:2023tsl,Buchmuller:2023aus,Figueroa:2023zhu,NANOGrav:2023hvm,EPTA:2023xxk}. With the latest NANOGrav 15-yr dataset, Ref.~\cite{Servant:2023mwt} put constraints on the axion-like particles (ALPs) mass ($m_a$) and the axion decay constant ($f_a$) based on semianalytic results of axion string~\cite{Gouttenoire:2019kij,Chang:2019mza,Chang:2021afa} and numerical results of domain walls after the QCD phase transition~\cite{Hiramatsu:2013qaa}.    

Axion and more general ALPs raise growing interest in different experimental searches~\cite{Irastorza:2018dyq,Arza:2019nta,DiLuzio:2020wdo,Sikivie:2020zpn,Mitridate:2020kly}, where the ALPs are motivated by many extensions of the Standard Model (SM) for rich phenomenology~\cite{Choi:2020rgn}. With the assistance of the interaction between axion/ALPs and Higgs, it was proposed to realize the electroweak symmetry breaking during the evolution of the Universe through the relaxation mechanism~\cite{Graham:2015cka,Espinosa:2015eda}. The interaction between the ALP and the SM Higgs leaves some imprints on the Cosmic Microwave Background (CMB)~\cite{Fung:2021wbz,Luu:2021yhl}.
 
When the axion coupled with the Higgs through the quartic coupling, there would emerge the Axion-Higgs string after the PQ symmetry breaking, where the non-winding classical Higgs configurations surrounding axion strings evolve and affect the CMB observation~\cite{Benabou:2023ghl}.   
In this work, we numerically study the properties of Axion string and Axion-Higgs strings with 3D lattice simulation in the early Universe. We show the log violation behavior of the scaling parameter, emission of the free axion, and the GWs for the Axion string and Axion-Higgs strings. Finally, we estimate the constraints on the axion decay constant $f_a$ and the axion mass $m_a$ with the DM relic abundance and dark radiation at the Big Bang nucleosynthesis (BBN). 

\section{The simulation setup}
We consider the Lagrangian density including a complex field respecting a global U(1) symmetry and a scalar Higgs field 
\begin{align}
	\mathcal{L} = -\partial_\mu \varphi^* \partial^\mu \varphi - \frac{1}{2} \partial_\mu s \partial^\mu s - V \left(\varphi, s, T\right) \label{eq:lagr}
\end{align}	
with $V \left(\varphi, s, T\right)=V_1\left(\varphi,T\right) +V_2\left(\varphi, s, T\right)$ and the two potentials of $V_{1,2}$ take the form of
\begin{align}
	V_1\left(\varphi,T\right) = & \lambda_\varphi\left(|\varphi|^2-\frac{f_a^2}{2}\right)^2+\left(\frac{\lambda_\varphi}{3} + \frac{\lambda_{\varphi s}}{6}\right) T^2|\varphi|^2 \;,\label{eq:p1} \\
	V_2\left(\varphi, s, T\right) = &\frac{1}{2} \gamma \left(T^2-T_0^2\right)s^2+\frac{1}{3}ATs^3+\frac{1}{4}\lambda_s s^4 + \\ \nonumber 
	&\frac{1}{2} \lambda_{\varphi s} |\varphi|^2 s^2\;. \label{eq:p2}
\end{align} 
As in Ref.~\cite{Hiramatsu_2011}, we included the thermal corrections to the effective potential of both $V_1$ and $V_2$, the thermal effective potential $V_1$ describes the broken of the global U(1) symmetry at low temperature, and $V_2$ is a general thermal potential of a scalar Higgs.
As indicated by Eq.\ref{eq:p1}, we consider the axion decay constant $f_a=v_a$, which means that we consider the domain wall number $N_{\rm DW}=1$ (corresponding with the KSVZ model~\cite{Kim:1979if,Shifman:1979if}).
We note that thermal corrections have been neglected in recent studies of Refs.~\cite{Benabou:2023ghl,Gorghetto_2018,Gorghetto:2020qws,Gorghetto:2021fsn,Baeza-Ballesteros:2023say}. And, different from Ref.~\cite{Benabou:2023ghl}, we here consider the system with ALP and a general scalar Higgs rather than specialized the system to QCD axion and the SM Higgs. The mass of the radial mode is therefore depends on the temperature as the Universe cools down: $m_r^2=2\lambda f_a^2+(\lambda_\phi/3+\lambda_{\phi s}/6)T^2$.

Utilizing the public code ${\mathcal CosmoLattice}$~\cite{Figueroa:2021yhd,Figueroa:2020rrl} and adopting the second-order leap-frog algorithm, we evolve the equation of motions (EOMs) from the Lagrangian Eq.~\ref{eq:lagr} in the radiation dominant universe. The lattice simulation takes grid number $N=1024$, and we use the conformal coordinate system $ds^2 =a^2(\eta)(-d\eta^2+d\mathbf{x}^2)$. We adopt dimensionless variables in the simulation, the field variables are scaled by $f_* = f_a$, and space-time variables are scaled by $\omega_* = a_i H_i$, where the subscript i denote initial time and the dimensionless conformal time is $\tilde \eta = a(\eta)/a_i$. The simulation begins when $T_i = 2T_c$, where $T_c=\sqrt{\lambda_\phi/(\lambda_\phi/3+\lambda_{\phi s})}f_a$ is the critical temperature of the U(1) symmetry breaking, and initial field and momentum configuration is given by thermal fluctuation (see details in Appendix.~\ref{app:lattice simulation}). For simplicity, We use $\lambda_\phi=1.0$. Considering the finite volume effect and discretization error, we set the initial physical box length $ L_i = 43.4(a_i H_i)^{-1}$ with $f_a = 2.54*10^{17}$ GeV, and evolve the EOMs from $\tilde{\eta_i}=1.0$ to $\tilde{\eta_e}=14.0$, which means the box contain 3.1 Hubble length per side and the cosmic string core width is two times lattice spacing at the end of the simulations. The time step is set to be $\delta \tilde{\eta}=0.01$, satisfying the Courant-Friedrichs-Lewy condition with $\delta \tilde{\eta} < \delta \tilde{x}/3 $.

\begin{figure}[!htp]
    \centering
\includegraphics[width=0.365\textwidth,clip]{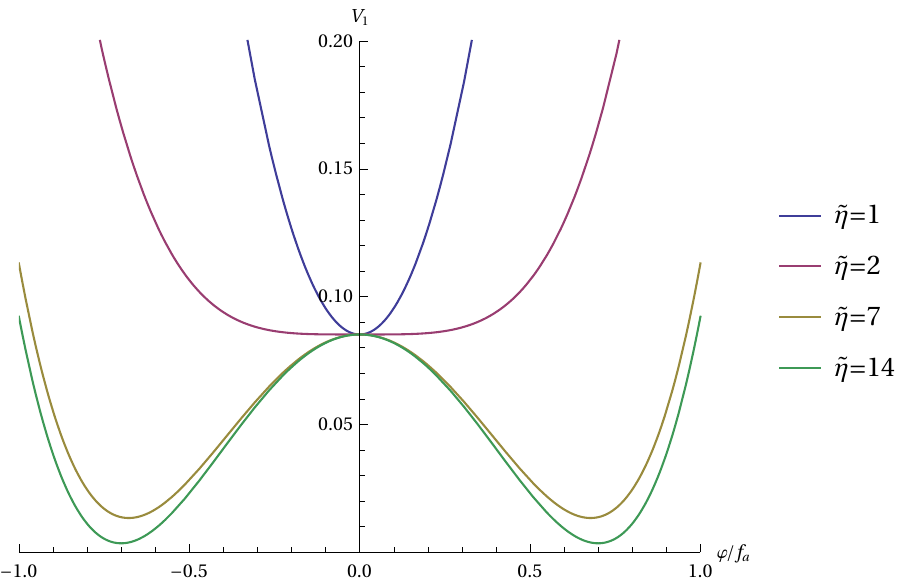}
    \centering
\includegraphics[width=0.365\textwidth,clip]{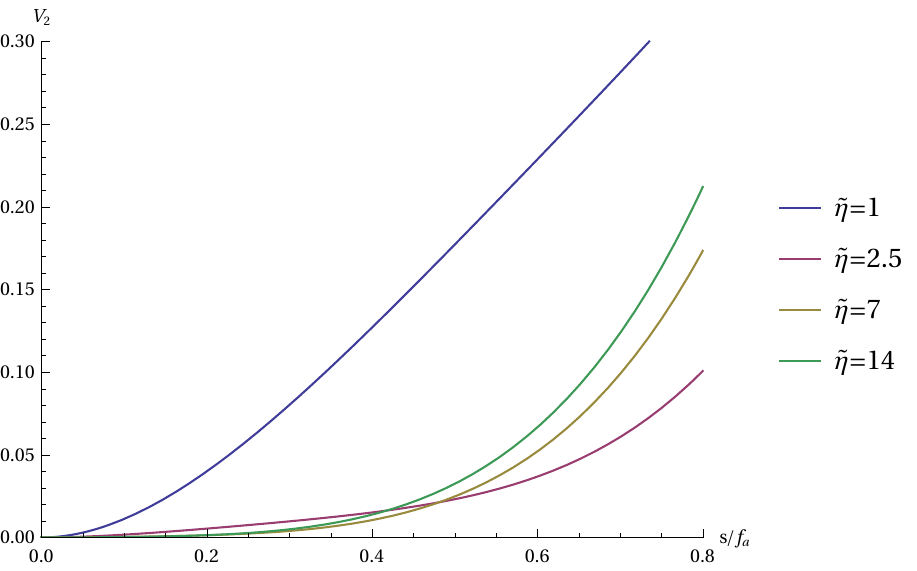}
    \centering
\includegraphics[width=0.365\textwidth,clip]{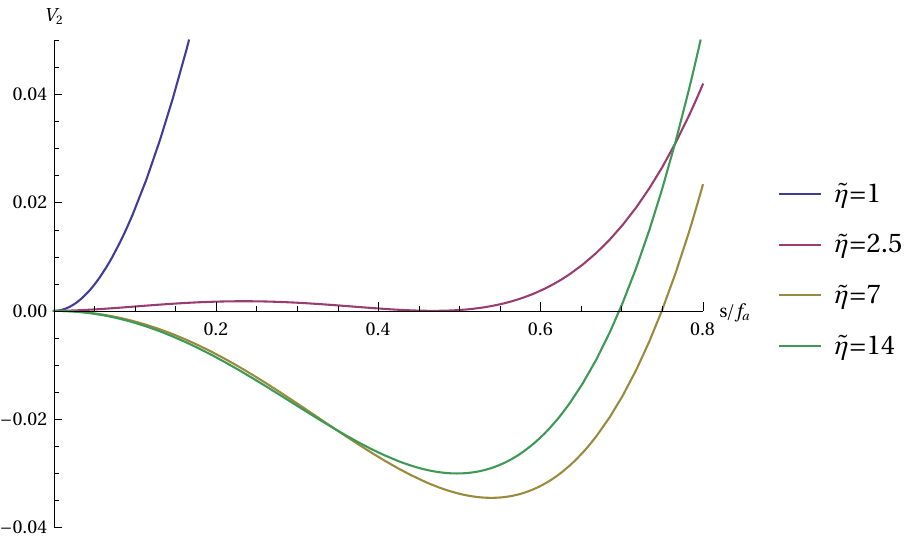}
    \caption{Potential shape for Axion-Higgs-1 at different times in the simulation. Top: $V_1$ with $s=0$. Middle: $V_2$ with $\varphi=\varphi_{vac}$. Bottom: $V_2$ with $\varphi=0$.}
    \label{fig:potential}
\end{figure}

We use parameters in Table.~\ref{tab:para} to ensure the production of the Axion-Higgs strings in the simulation. We also perform the simulation of pure axion string without the Higgs field by taking $\lambda_{\varphi s} = 0$. As an illustration, we show the potential shape for the Axion-Higgs-1 case at different time slices in Fig.\ref{fig:potential}. Where, the scalar Higgs $s$ never gets vacuum expectation value (VEV) at the places where PQ symmetry breaking occurs with $\varphi=v_a$, and the $s$ get VEV only at the places where PQ symmetry didn't get broken with $\varphi=0$. 

\begin{table}[!htp]
    \centering
    \begin{tabular}{c|ccccc}
        & $\gamma$ & $T_0$ & A & $\lambda_s$ & $\lambda_{\phi s}$ \\
        \hline     
    Axion-Higgs-1 &  2.75 & $1.78*10^{17}$ &-5.57 & 6.96 & 3.0\\
    Axion-Higgs-2 &  2.35 & $1.78*10^{17}$ &-5.81 & 8.87 & 3.0\\
    \end{tabular}
    \caption{Benchmark parameters in the thermal potential (Eqs.~(\ref{eq:p1},\ref{eq:p2})) for the simulations of the Axion-Higgs string systems.}
    \label{tab:para}
\end{table}

\section{Numerical results}
In this section, we first give the justification for the existence of the Axion-Higgs strings through numerical estimation of the field configurations.   
We use the dimensionless variables to calculate the Axion-Higgs profiles, where we introduce the notation $\varphi/f_*=g\left(r\right)\exp{i\theta}$, $s/f_*=s\left(r\right)$ and $r_{phy}\omega_*=r$ in cylindrical coordinate for infinitely straight string Higgs (string) along the z direction. The EOMs cast the form of
\begin{align}
    g^{\prime\prime}(r) + &\frac{1}{r}g^{\prime}(r)-\frac{1}{r^2}g(r) = \nonumber \\ &\lambda_\varphi \frac{f_*^2}{\omega_*^2} g(g^2-\frac{1}{2})+(\frac{\lambda_\varphi}{3}+\frac{\lambda_{\varphi s}}{6})\frac{T^2}{\omega_*^2}g+\frac{\lambda_{\phi s}}{2}\frac{f_*^2}{\omega_*^2}gs^2\;, \\
    s^{\prime\prime}(r) + &\frac{1}{r}s^{\prime}(r) = \nonumber \\ &\lambda_{\phi s}\frac{f_*^2}{\omega_*^2}g^2s+\gamma \frac{T^2-T_0^2}{\omega_*^2}s+AT\frac{f_*}{\omega_*^2}s^2+\lambda_s\frac{f_*^2}{\omega_*^2}s^3\;\nonumber.
\end{align}
Note that $^\prime = \frac{\partial}{\partial r}$ in above equations. With the boundary conditions $g(0)=0, s'(0)=0, g(\infty)=1/\sqrt{2}, s(\infty)=0$, we can numerically solve the profile of Axion-Higgs strings. 

\begin{figure}[!htp]
    \centering
    \includegraphics[width=0.4\textwidth,clip]{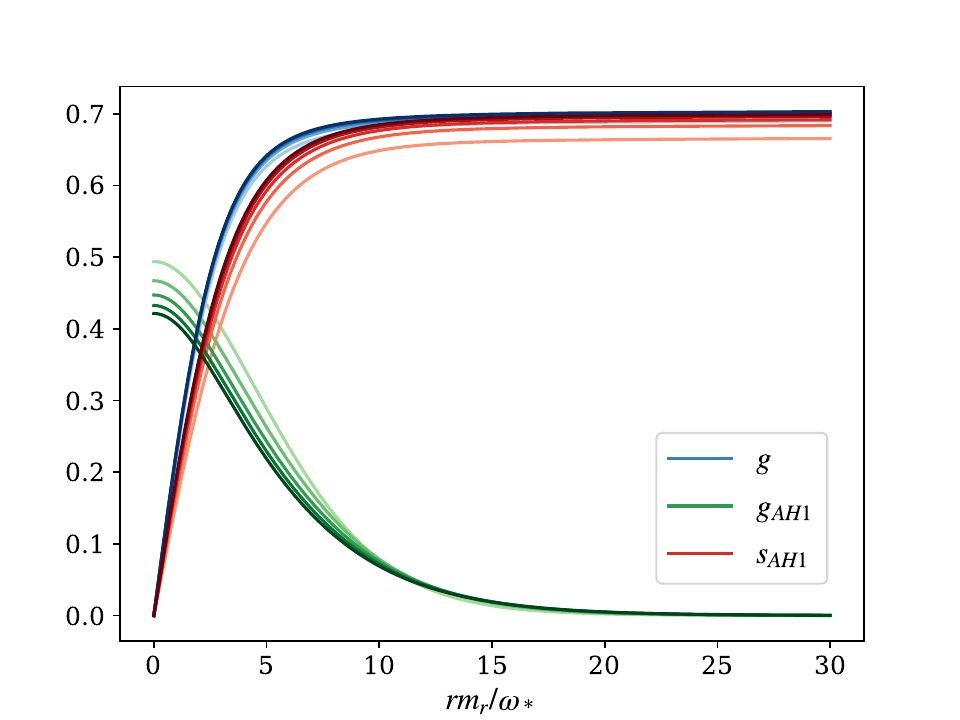}
    \centering
    \includegraphics[width=0.4\textwidth,clip]{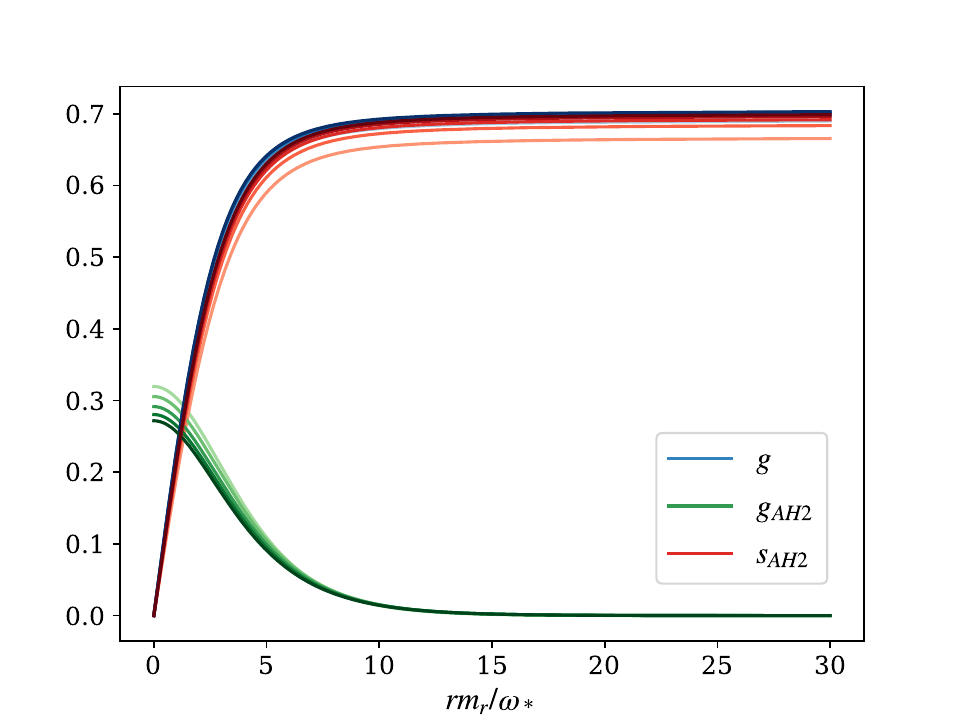}
    \caption{Field value distribution along the radial direction for Axion-Higgs-1 (with $g_{AH1}$ and $s_{AH1}$) and Axion-Higgs-2 (with $g_{AH1}$ and $s_{AH1}$). Deeper color means increasing time from $\tilde{\eta}=6.0$ with interval $\delta \tilde{\eta}=2.0$. The blue line is the solution of the Axion string.}
    \label{fig:profile}
\end{figure}

\begin{figure}[!htp]
    \hfill
\includegraphics[width=0.2\textwidth,clip]{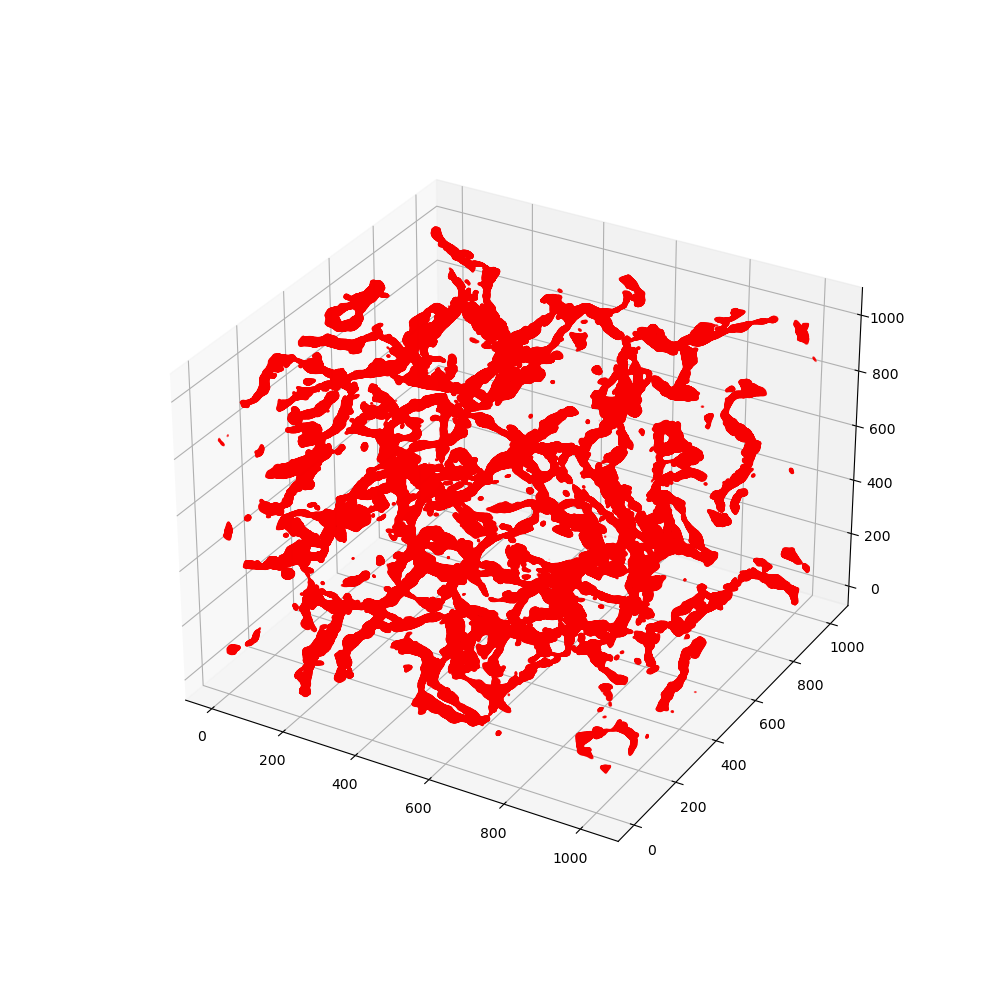}
    \hfill
    \includegraphics[width=0.2\textwidth,clip]{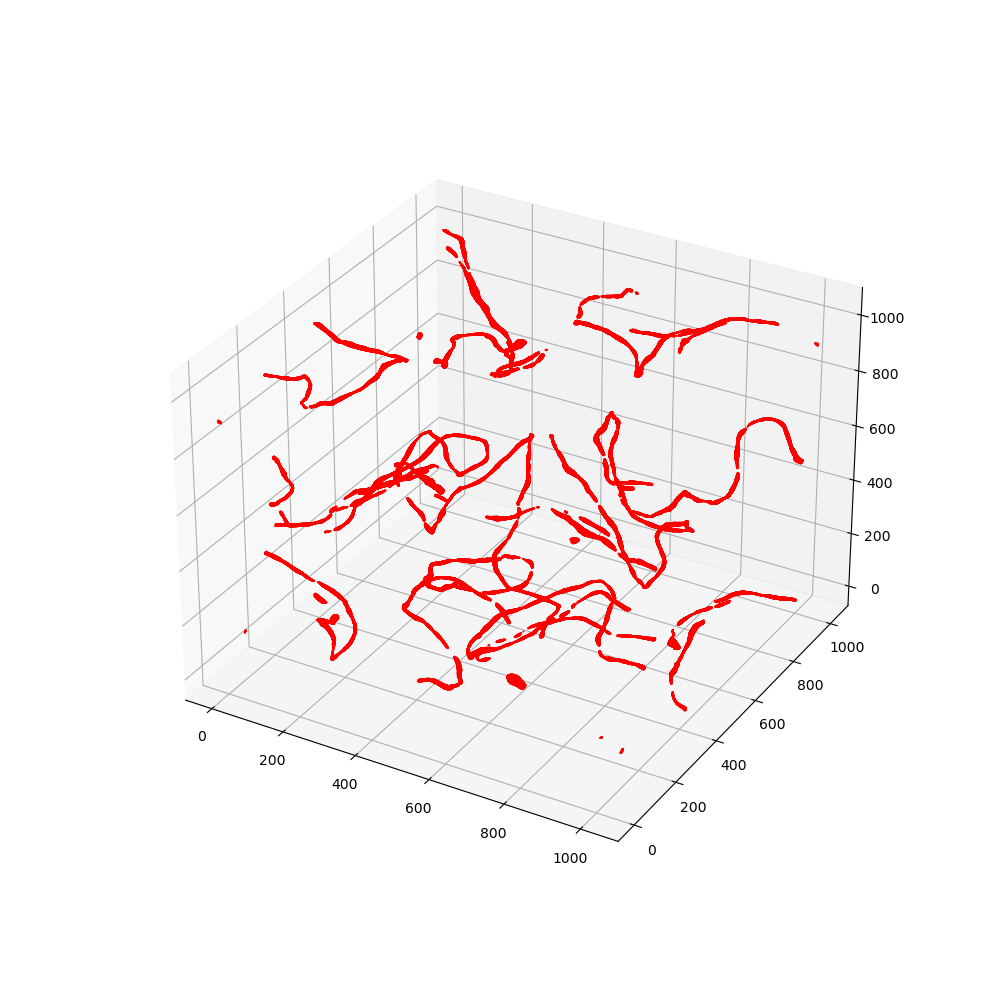}
    \\ 
    \hfill
    \includegraphics[width=0.2\textwidth,clip]{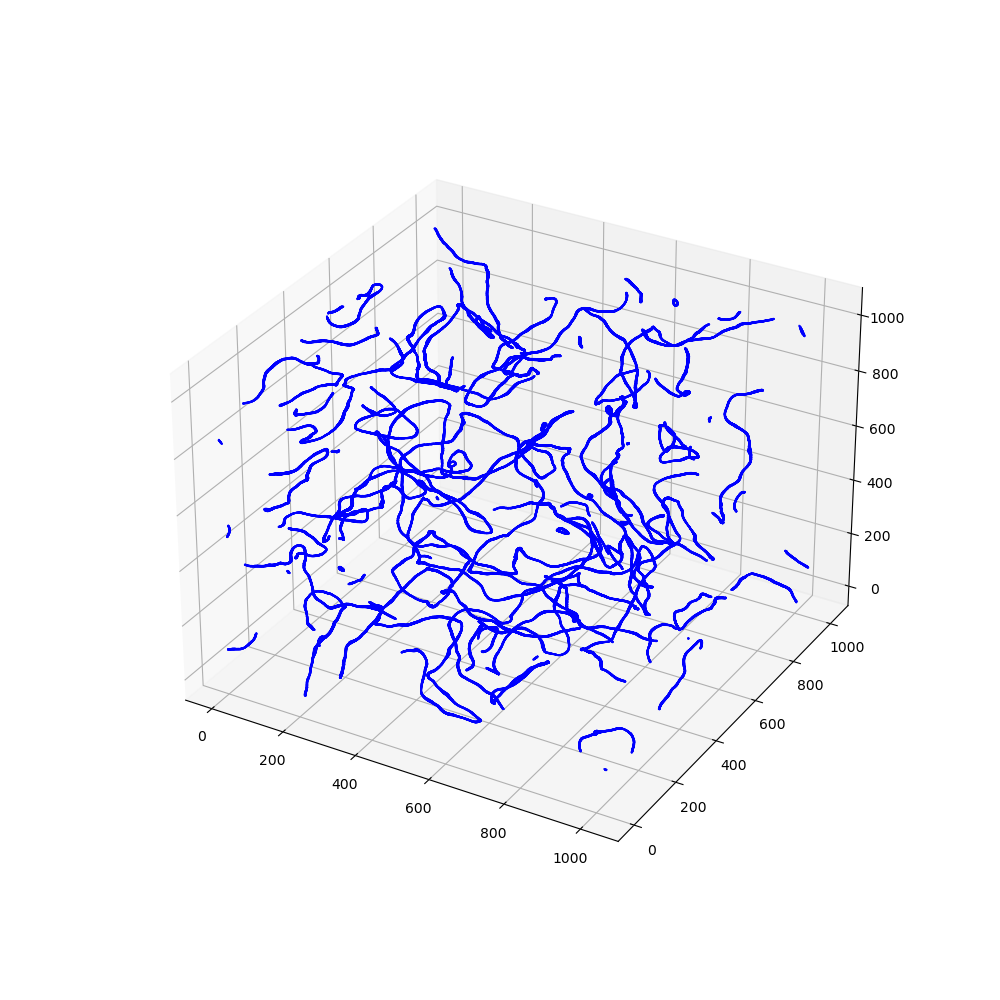}
    \hfill
    \includegraphics[width=0.2\textwidth,clip]{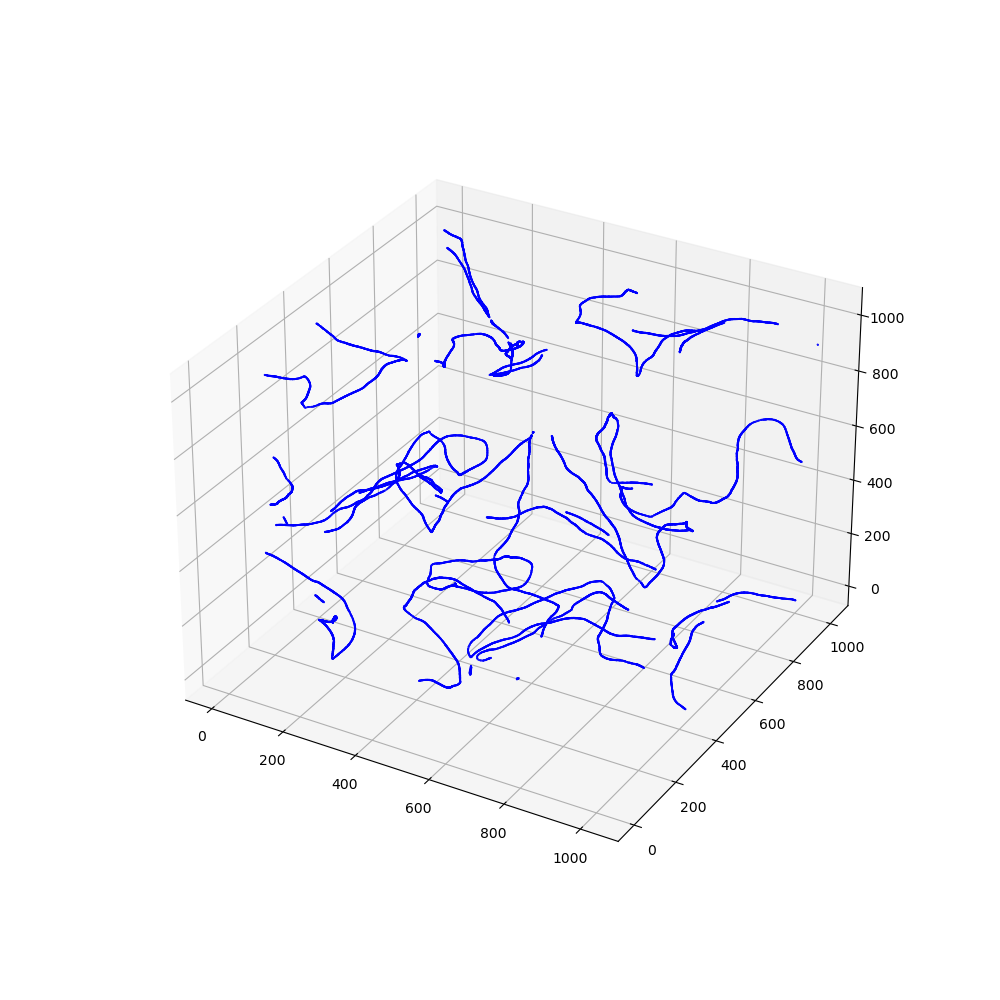}
    \caption{3D snapshot for complex field scalar field $ s>0.35 f_a$ (top) and  $|\varphi|<0.1 f_a$ (bottom) at the times of $\tilde{\eta}=9.0,14.0$ in the Axion-Higgs-1 simulation (from left to right).} 
    \label{fig:snapshot}
\end{figure}

In Fig.\ref{fig:profile}, we present the Axion-Higgs strings profile of the two benchmarks given in Table~\ref{tab:para} at different times. The two plots show that the scalar fields have different values in the string core for the two Axion-Higgs string cases and change little with time evolution. For illustration, we show the snapshot of field configurations in Fig.\ref{fig:snapshot} for one benchmark, which clearly demonstrates the axion and Higgs fields evolve in the simulation with analogous configurations.

The string properties can be captured with the string tension and scaling parameters during the simulations. We first study the tensions of the axion string and the Axion-Higgs strings.
The dimensionless tension can be theoretically obtained after one gets the axion string and Axion-Higgs strings through $\tilde{\mu} = \mu/f_*^2=\int \tilde{\rho} r\mathrm{d}r\mathrm{d}\theta$. More exactly,  
\begin{align}
        \tilde{\mu} 
        &= 2\pi \int \left((g'^2+\frac{g^2}{r^2} + (s'^2 +\tilde{V}(g,s) \right) r \mathrm{d}r\;.
\end{align}

\begin{figure}[!htp]
    \centering
     \includegraphics[width=0.4\textwidth,clip]{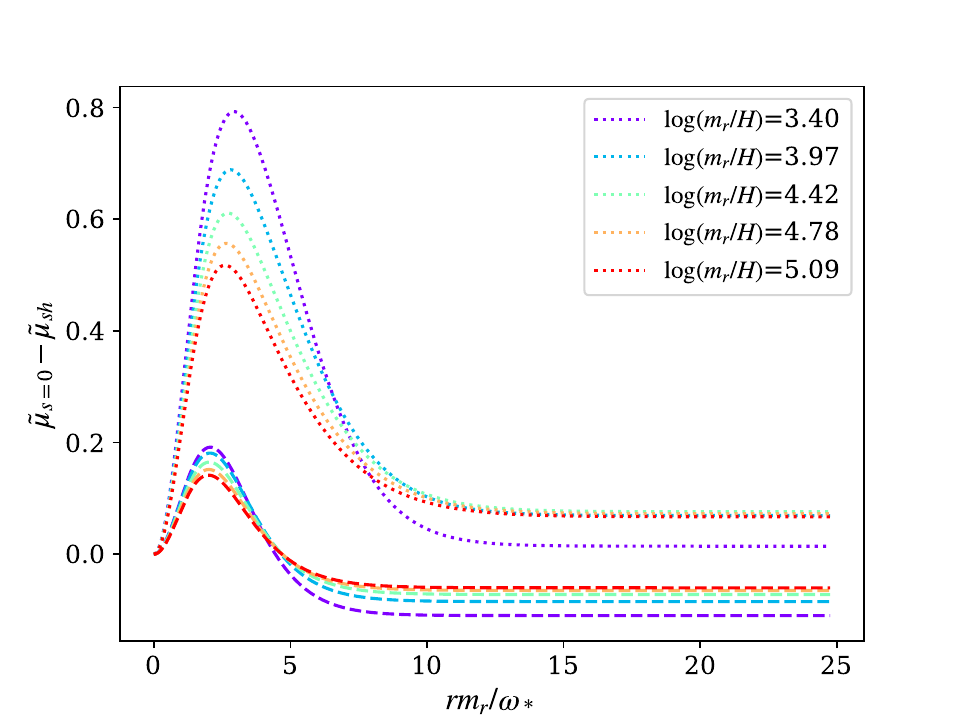}
    \includegraphics[width=0.4\textwidth,clip]{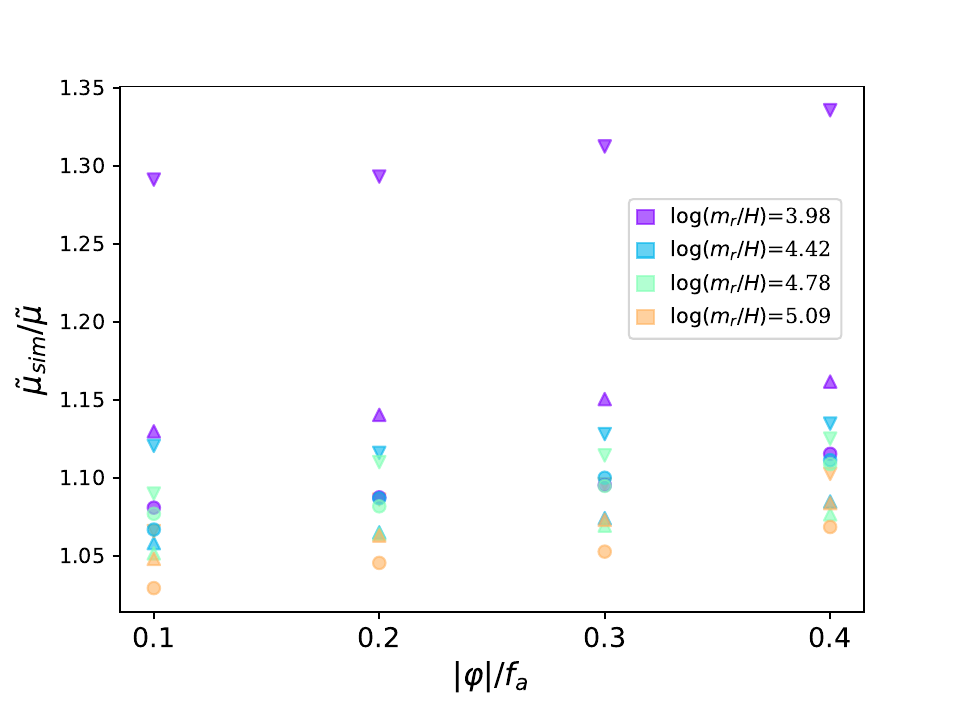}
     \caption{Top: The difference between string tensions of Axion-Higgs strings and Axion string along the distance from string core, where the dotted lines are the result of Axion-Higgs-1 and the dashed lines are the result of Axion-Higgs-2; 
    Bottom: the ratios of the string tension obtained through simulations and theoretical calculations for Axion-Higgs-1 (upperward triangles), Axion-Higgs-2 (downward triangles) and axion (circle dots).
    \label{fig:tension}}
\end{figure}

We first show the tension difference in the top panel of Fig.\ref{fig:tension} for the two Axion-Higgs string cases. Both the case of Axion-Higgs-1 and Axion-Higgs-2 give the lower tension compared with the solution with $s=0$ around the string cores, this property ensures the stability of the Axion-Higgs strings in the simulations. Though our thermal potential is different from Ref.~\cite{Benabou:2023ghl}, we also observe that the scalar Higgs contributes to the string tension at the order of $f_a^2$. 
The Axion-Higgs string here under study is different from the electroweak axion strings proposed by Ref.~\cite{Abe:2020ure} where the Higgs exhibits non-trivial winding.
The string tension can be calculated by $\tilde{\mu}_{sim} = \rho_s / (4 \xi H^2 f_a^2) $ through the simulation with $ \rho_s $ being the energy density of the $|\varphi| / f_a < const$ area. In the bottom plot, we show the contrast between theoretical calculations and simulations with different criteria (upper limit of complex field value). The results fit better as time evolves for all three cases. 

\begin{figure}[!htp]
    \centering
    \includegraphics[width=0.4\textwidth,clip]{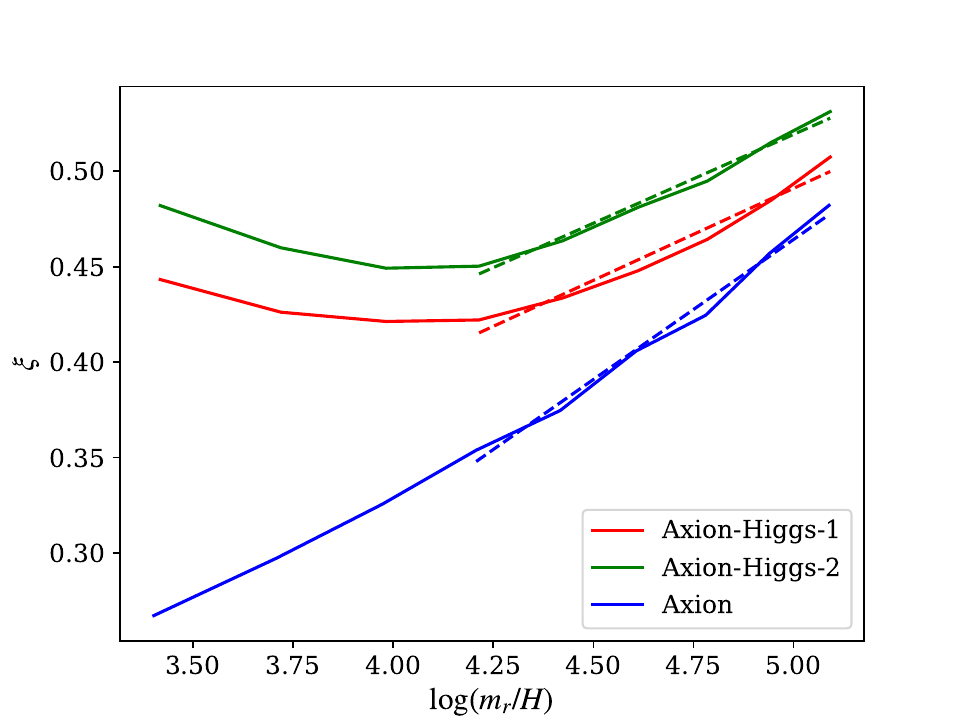}
    \caption{Scaling parameter evolution in the simulations (solid lines) and the fitting results (dashed lines).}
    \label{fig:scalingpara}
\end{figure}

 We show the scaling parameter of the axion string and the Axion-Higgs strings cases in Fig.\ref{fig:scalingpara}, where all the scaling parameters enter the linear growth of $\log(m_r/H)$ as observed in Refs.~\cite{Buschmann_2022,Gorghetto_2018}. The detailed calculation of the scaling parameters for the string is given by Appendix.~\ref{app:scaling parameter}. We take the fit function $\xi = \alpha \log(m_r/H) + \beta$, and get the value of $\alpha=0.096, 0.093, 0.15$ and $\beta=-0.009, 0.054, -0.27$ for Axion-Higgs-1, Axion-Higgs-2 and axion string cases, respectively.

\begin{figure}[!htp]
    \centering
    \includegraphics[width=0.4\textwidth,clip]{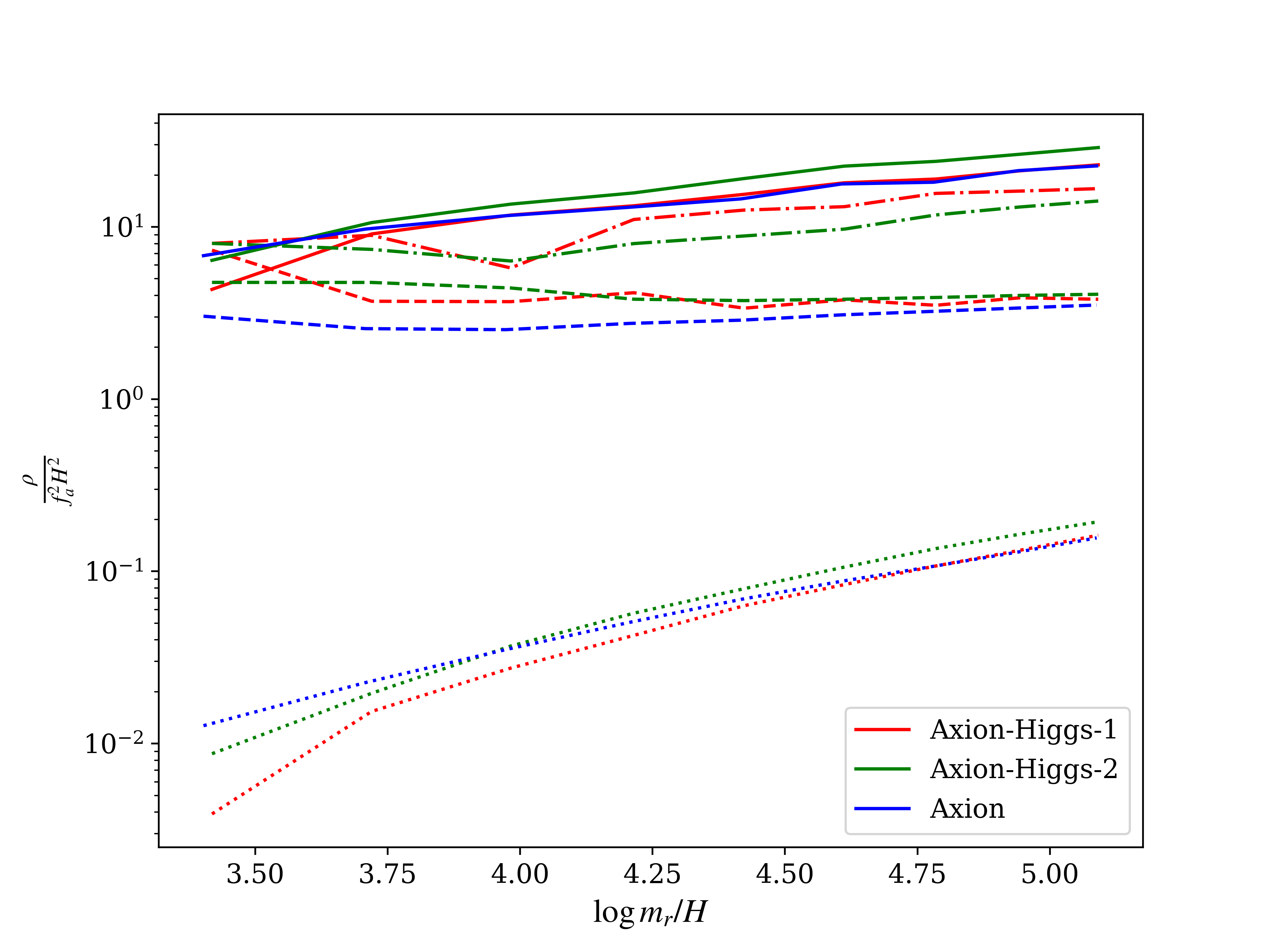}
    \caption{Reduced energy density of free axions, Higgs fields (dash-dotted line), strings (dashed line) and GWs (dotted line)}.\label{asgw} 
\end{figure}

We then show the energy density of free axion, string, gravitational wave, and Higgs field in Fig.\ref{asgw}, and the string energy is calculated by the energy density of area $\phi/f_a<0.4$, see the Appendix for more details on the calculations of different energy density components. We observe that the ratio between the axion particles and GWs radiation is around $\rho_{gw}/\rho_{a}\sim\mathcal{O}(10^{-3}-10^{-2})$, which indicates that the free axions emitted from strings dominate over GWs radiation.

\begin{figure}[!htp]
    \centering
    \includegraphics[width=0.4\textwidth,clip]{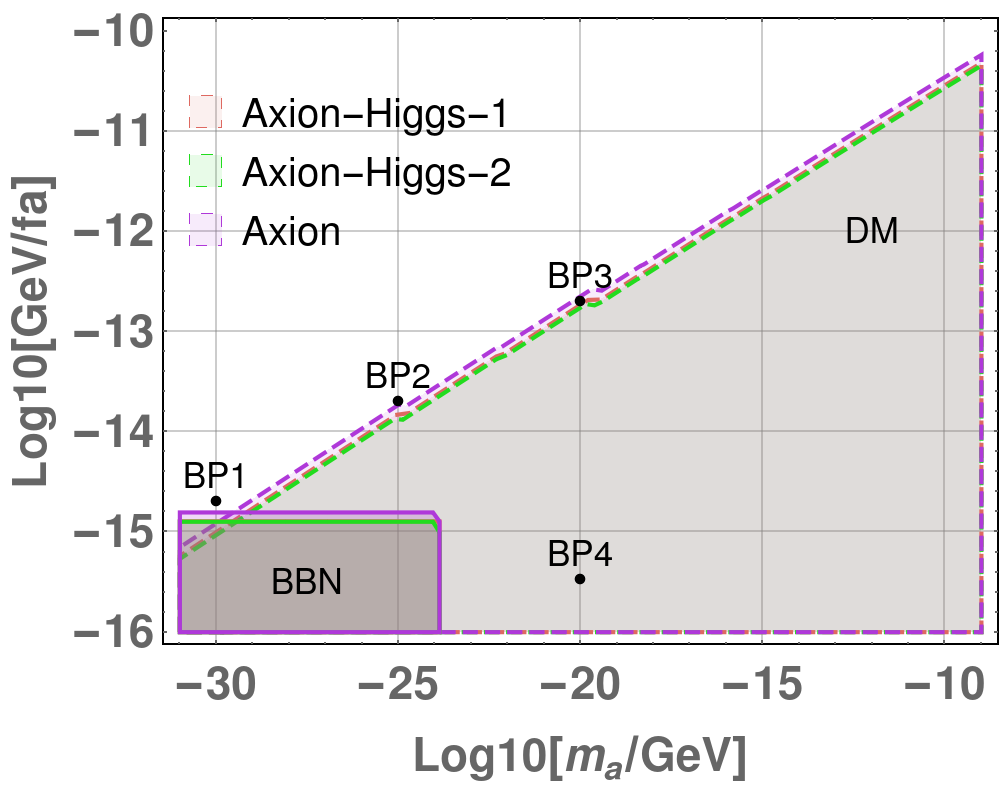}
    \caption{constrain for $U(1)$ symmetry breaking energy scale $f_a$ and mass of ALP $m_a$ from dark matter and dark radiation at BBN. }
    \label{fig:constrainALP}
\end{figure}

Assuming that the loss of string energy is all converted into free axion and the mean reciprocal comoving momentum of radiated axion $ k^{-1}(t) \propto t $, we can integrate the total number density of axion from string formation to QCD scale and redshift it to today, which can be regarded as DM (for details see Appendix.\ref{app:axion spectrum}). Considering the constraints from the cold DM relic abundance $ \Omega_{DM}h^2<0.12 $, we get the upper limit for the PQ breaking scales: $f_a < 5.1*10^9, 5.6*10^9, 3.7*10^9 \rm GeV$ respectively. For the case of ALPs radiated from the strings, the difference compared with QCD axion calculation is that the mass of ALPs is a free parameter, which changes the end time of the axion emission from strings determined by the mass. The ALP DM relic density is then given by  
\begin{align}
    \Omega_{DM} h^2 \approx 1.6 \times 10^9 &f_a^2 m_a^2 \xi (-2.25+\log{\frac{f_a}{m_a\xi^{1/2}}}) \nonumber \\ &\times\epsilon^{-1} (m_a^2 m_{pl}^2 g_*^{1/3})^{-\frac{3}{4}}\;.
\end{align}
Where, $m_{pl} = 1/ \sqrt{8\pi G}=2.44\times10^{18} {\rm GeV}$ is the reduced Planck mass, $g_*$ and $\xi$ is the effective number of relativistic degrees of freedom and scaling parameter at the end of simulations, the dimensionless parameter $\epsilon^{-1}$ is extracted from the simulations following the method given in Ref.~\cite{Hiramatsu_2011}, and $\epsilon^{-1}=0.87, 0.78, 0.85$ for Axion-Higgs-1, Axion-Higgs-2, and Axion strings.
We also calculate the energy density of the free axion of the ALPs at the BBN time $t_{BBN}$ that contribute to the effective number of neutrinos as $\Delta N_{eff} = (8/7)(11/4)^{4/3} \rho_a/\rho_\gamma$ at temperature of $1 \text{MeV}$, which is given by 
\begin{align}
    \Delta N_{eff} &= \xi(\frac{f_a}{10^{15} {\rm GeV}})^2  \bigg(-1.21\times10^{-3} \nonumber  \\
&+1.67\times10^{-5}\log^2\big(\frac{1.12\times10^{39} \frac{f_a}{10^{15} {\rm GeV}} }{\xi^{1/2}}\big)\bigg)\;.
\end{align}
The conservative upper bound of dark radiation from BBN is $\Delta N_{eff} < 0.46$ \cite{Planck:2018vyg}. 
We show the constraint for ALPs parameter space from DM overabundance and dark radiation limitation from BBN in Fig.\ref{fig:constrainALP}. ALPs' evolution might also have complicated implications for CMB detection, which is beyond the scope of this work, and we leave it for future study. 

\begin{figure}
    \centering
    \includegraphics[width=0.5\textwidth,clip]{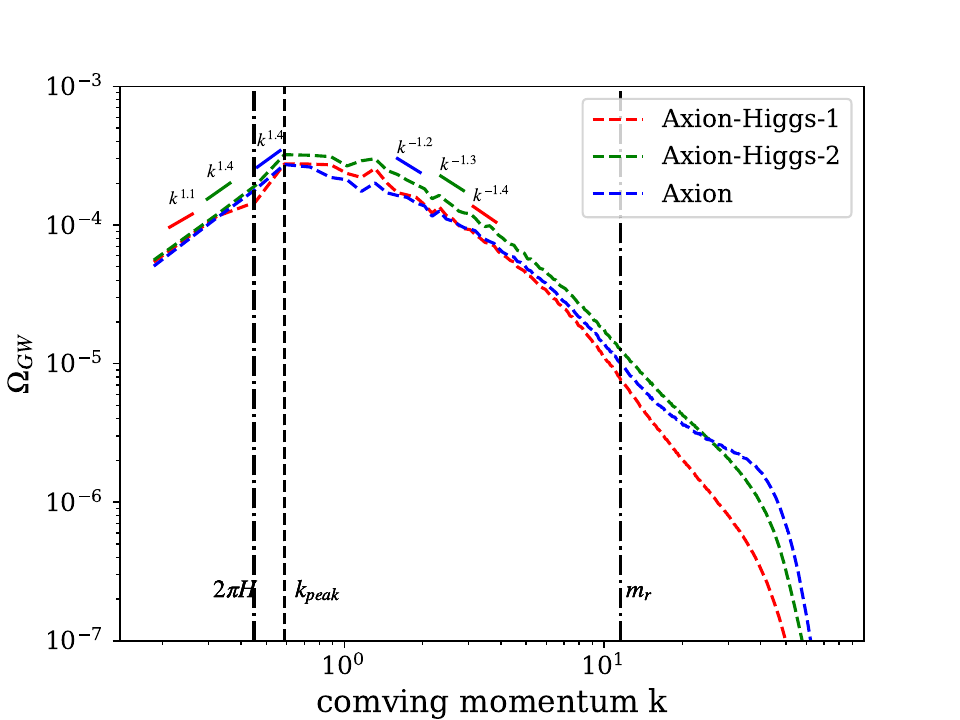}
    \caption{GWs spectra at the end of simulations ($\log{m_r/H}=5.09$).} \label{fig:gwspectra}
\end{figure}

\begin{figure}
    \centering
    \includegraphics[width=0.5\textwidth,clip]{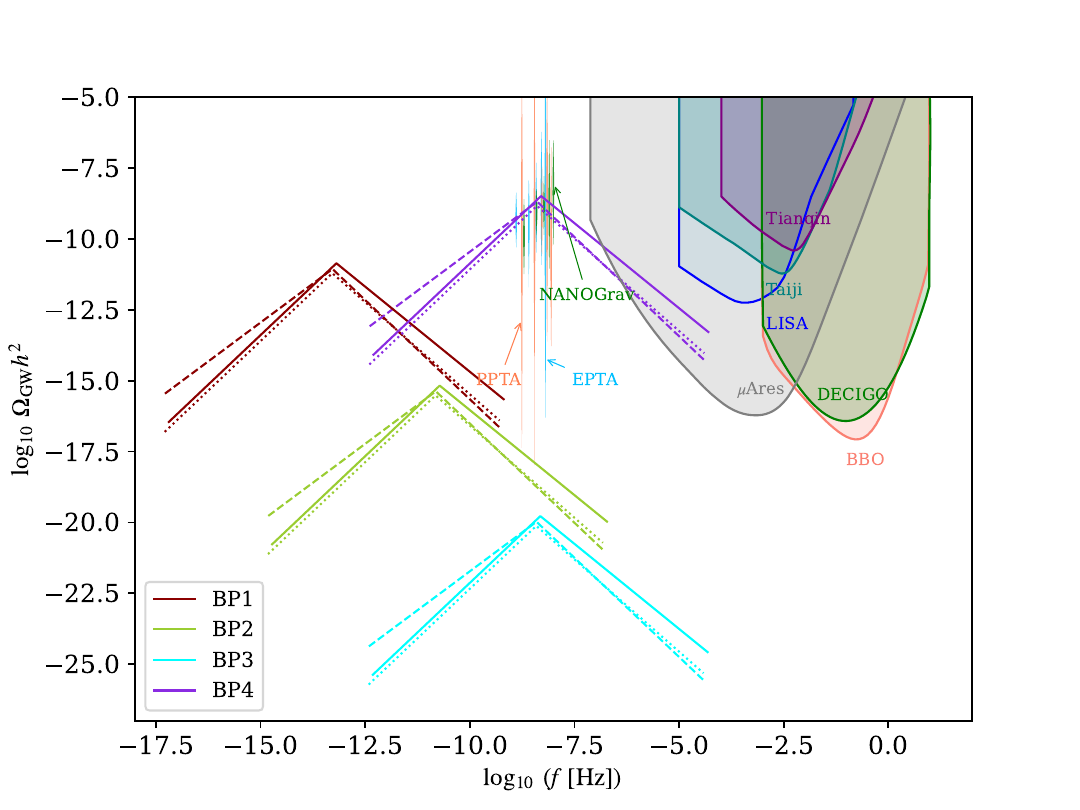}
    \caption{The GW spectra today for the three benchmark points for Axion-Higgs-1(dashed line), Axion-Higgs-2(dotted line) and axion(solid line) given in Fig.\ref{fig:constrainALP}.}
    \label{fig:gwtoday}
\end{figure}

Following the conventional GWs computation method through the lattice simulation (detailed in Appendix.\ref{app:gws power spectrum}), we obtain the GW spectra at the end time of simulations, see Fig.\ref{fig:gwspectra}.
Then $\rho_{gw}$ is calculated by integrating the instantaneous emission $\mathrm{d}(a^4 \rho_{gw} / \mathrm{d}t)$, which is given by $\mathrm{(d(a^4\rho_{gw})/dt} / \mathrm{ (d(a^4\rho_a)/dt)} = \gamma G \mu^2/f_a^2$ \cite{Gorghetto:2021fsn}, from string formation to the end, and we get the value of $\gamma = 0.03, 0.02, 0.03$ for the three cases of Axion-Higgs-1, Axion-Higgs-2, and Axion string. All three cases yield similar GW spectra with $\Omega_{GW}\propto k^{1(-1)}$ on the left (right) of the peak frequency. Assuming the shape of the GWs power spectra is unchanged after the end of the simulations and the frequency of peak of the power spectrum $\sqrt{\xi}H$ corresponds to the average distance between long strings, we use $\rho_{gw}$ to reconstruct the GWs power spectra at the end of string evolution and redshift the spectrum to today. We take the benchmark points satisfying the DM and dark radiation constraints given in Fig.\ref{fig:constrainALP}(except BP4) and show the GWs power spectra in Fig.~\ref{fig:gwtoday}. Our results show that all the Axion and Axion-Higgs strings (BP1,2,3) cannot be probed by current PTA experiments and future space-based GW detectors, such as LISA ~\cite{LISA:2017pwj,Baker:2019nia}, TianQin~\cite{TianQin:2015yph,Zhou:2023rop}, Taiji~\cite{Hu:2017mde,Ruan:2018tsw}, $\mu Ares$~\cite{Sesana:2019vho}, DECIGO~\cite{Seto:2001qf,Kawamura:2011zz,Yagi:2011wg,Isoyama:2018rjb}, and BBO~\cite{Crowder:2005nr,Corbin:2005ny,Harry:2006fi}, and the GWs power spectra can be detected by PTA experiment is excluded by the ALPs dark matter constraint (see BP4).  

We find that Ref.\cite{Gorghetto_2021} gives slightly tighter restrictions for $f_a$ and $m_a$ in comparison with our work from DM overproduction and is almost the same limitation from BBN constraint. The GWs spectra in Fig.~\ref{fig:gwtoday} are steeper and the peak amplitude is higher than Ref.\cite{Gorghetto_2021} by one order of magnitude, and the peak frequency is smaller than Ref.\cite{Gorghetto_2021} by around one-to-two order. Note that Ref.\cite{Gorghetto_2021} uses the method of integrating the instantaneous emission spectrum to calculate the axion and GWs spectra to get a smoother slope on the right-hand side of the peak frequency.

\section{Conclusions and discussions}

We study the cosmological implication of Axion string and Axion-Higgs strings through 3D lattice field simulations. In comparison with the Axion string scenario, the Axion-Higgs string scenarios have slightly larger string tensions and scaling parameters. However, we observe only moderate difference for free axion and gravitation wave radiation from Axion-Higgs strings and axion strings, and for these scenarios, the axion radiations dominated over that of gravitational wave, with $\rho_{gw}/\rho_a\sim \mathcal{O}(10^{-3}-10^{-2})$. Considering the constraints from DM relic abundance and the BBN bound on dark radiations from the relativistic axion and ALPs, we found the GWs from purely axion strings and Axion-Higgs strings cannot be probed by the current datasets of PTA experiments (EPTA, PPTA, NANOGrav, and CPTA), and the Axion string and Axion-Higgs string scenarios that can interpret the evidence of stochastic GW backgrounds in the dataset of the latest PTA experiments have been excluded by the DM relic abundance, this is consistent with the observation of Ref.~\cite{Servant:2023mwt}. Therefore we conclude that to explain the possible signature of stochastic GWs observed by PTAs, one needs to consider local string rather than global string.

We only study the particle and GWs radiations from cosmic strings through the PQ era simulations in this work. For the numerical simulations of the cosmological evolution of axions in both the PQ era and QCD era where one has axions and GWs decay from domain walls, we refer to Refs.~\cite{Li:2023gil} for the case of domain wall number $N_{\rm DW}>1$ (the so-called DFSZ model~\cite{Dine:1981rt,Zhitnitsky:1980tq}) and Ref.~\cite{Buschmann:2019icd} for the case $N_{\rm DW}=1$ (the KSVZ model~\cite{Kim:1979if,Shifman:1979if}).    
 
\section{Acknowledgements.}
The numerical calculations in this study were carried out on the ORISE Supercomputer.
This work is supported by the National Key Research and Development Program of China under Grant No. 2021YFC2203004.
L.B. is supported by the National Natural Science Foundation of China (NSFC) under Grants Nos. 12075041, 12322505, and 12347101.
L.B. also acknowledges Chongqing Natural Science Foundation under Grant
No. CSTB2024NSCQ-JQX0022 and 
Chongqing Talents: Exceptional Young Talents Project No. cstc2024ycjh-bgzxm0020.

\appendix

\section{Lattice Simulation}
\label{app:lattice simulation}
We give a detailed realization of the discretization scheme in this section.
We take the following action  
\begin{align}
    S =& - \int \mathrm{d} x^4 \sqrt{-g}
    \left( \partial_\mu \varphi^* \partial^\mu \varphi + \frac{1}{2} \partial_\mu s \partial^\mu s+ V\left(\varphi, s, T\right)\right) 
\end{align}
with $V\left(\varphi, s, T\right)= V_1\left(\varphi,T\right) + V_2\left(\varphi, s, T\right)$ to describe the Axion-Higgs interaction in our simulation. The convention of FLRW metric is $ds^2 =a^2(\eta)(-d\eta^2+d\mathbf{x}^2)$.
Physical variables are converted into dimensionless ones through
\begin{align}
    \mathrm{d}\tilde{\mathbf{x}} = \omega_* \mathrm{d}\mathbf{x} \quad &\mathrm{d}\tilde{\tau} = \omega_* \mathrm{d}\tau \quad \tilde{\varphi} = \frac{\varphi}{f_*} \quad \tilde{s} = \frac{s}{f_*} \quad \;,\nonumber \\ 
    &\tilde{V} = \frac{V\left(f_* \tilde{\varphi}, ,f_* \tilde{s}, T\right)}{f_*^2\omega_*^2} \;,
\end{align}
and the equation of motions in this theme is  
\begin{align}\label{eoms}
    \tilde{\varphi}^{\prime \prime} - \tilde{\nabla}^2 \tilde{\varphi} + 2\frac{a^{\prime}}{a} \tilde{\varphi}^{\prime} &= -a^2\frac{\partial \tilde{V}}{\partial|\tilde{\varphi|}}\frac{\tilde{\varphi}}{2|\tilde{\varphi|}} \;, \\ 
    \tilde{s}^{\prime \prime} - \tilde{\nabla}^2 \tilde{s} + 2\frac{a^\prime}{a} \tilde{s}^{\prime} &= -a^2 \frac{\partial \tilde{V} } {\partial \tilde{s}}\;,
\end{align}
where $ \tilde{V} = \tilde{V_1} + \tilde{V_2}$ and $ ^\prime = \partial / \partial \tilde{\eta}$ is the time derivative. Solving the Friedmann equation with a radiation-dominated background, We have a relation 
\begin{equation}
    \tilde{\tau} = \frac{\tau}{\tau_i} = \frac{a}{a_i} = \left(\frac{T}{T_i}\right)^{-1} = \left(\frac{H}{H_i}\right)^{-\frac{1}{2}} \;.
\end{equation}

We use the thermal fluctuation as the initial condition of the simulation. The complex field is normalized by $\varphi = \frac{\phi_1+i\phi_2}{\sqrt{2}}$. The power spectrum of $\phi_i$ in momentum space is initialized by
\begin{equation}
    \mathcal{P}_{\phi_i}(k) = \frac{1}{\omega_k (e^{\omega_k/T} - 1)} \quad 
    \mathcal{P}_{\pi_{\phi_i}}(k) = \frac{\omega_k}{ e^{\omega_k/T} -1} \;.
\end{equation}
where $\pi_{\phi_i}$ is the time derivative of $\varphi_i$, $\omega_k = \sqrt{k^2/a^2 + m^2}$ and $ m_\varphi^2 = (\lambda_\varphi/3 + \lambda_{\phi s}/6)T^2 - \lambda_\phi v_\phi^2$ is the squared mass of complex field. The scalar field is also the thermal fluctuation with $m_s^2 = \gamma \left(T^2-T^2_0\right)$. The initial field configuration is given by Fourier transform in momentum space. 

\section{Scaling parameter of cosmic string}
\label{app:scaling parameter}
We use the phase of the complex scalar field $\theta = a/f_a$ to identify cosmic strings like the method of \cite{Hiramatsu_2011}. The plaquettes are pierced by string is identified by the minimum phase range $\delta \theta$ containing all four vertices exceeds $\pi$. Ordering the four phase in the plaquette with $\theta_1 > \theta_2 > \theta_3 > \theta_4$, and $\delta \theta$ is chosen by the minimum one of $ \left[ \theta_4-\theta_1, 2\pi-(\theta_2-\theta_1), 2\pi-(\theta_3-\theta_2), 2\pi-(\theta_4 - \theta_3) \right]$. We calculate the total number of the penetrated plaquettes in all three directions and multiply it by the comoving box spacing to get the comoving length of string $l_s$, where extra factor 2/3 due to the Manhattan effect \cite{Fleury_2016} is introduced. The number density of string per Hubble patch is defined as
\begin{align}
    \xi(t) = \lim\limits_{L \to \infty} \frac{L(t)t^2}{L(t)^3} = \frac{l_s t^2}{R(t)^2 V} \;,
\end{align}
where $V$ is the comoving box volume and t is given by $t = \frac{1}{2H}$. 

\section{Axion spectrum}
\label{app:axion spectrum}
\begin{figure}[!htp]
    \centering
    \includegraphics[width=0.4\textwidth,clip]{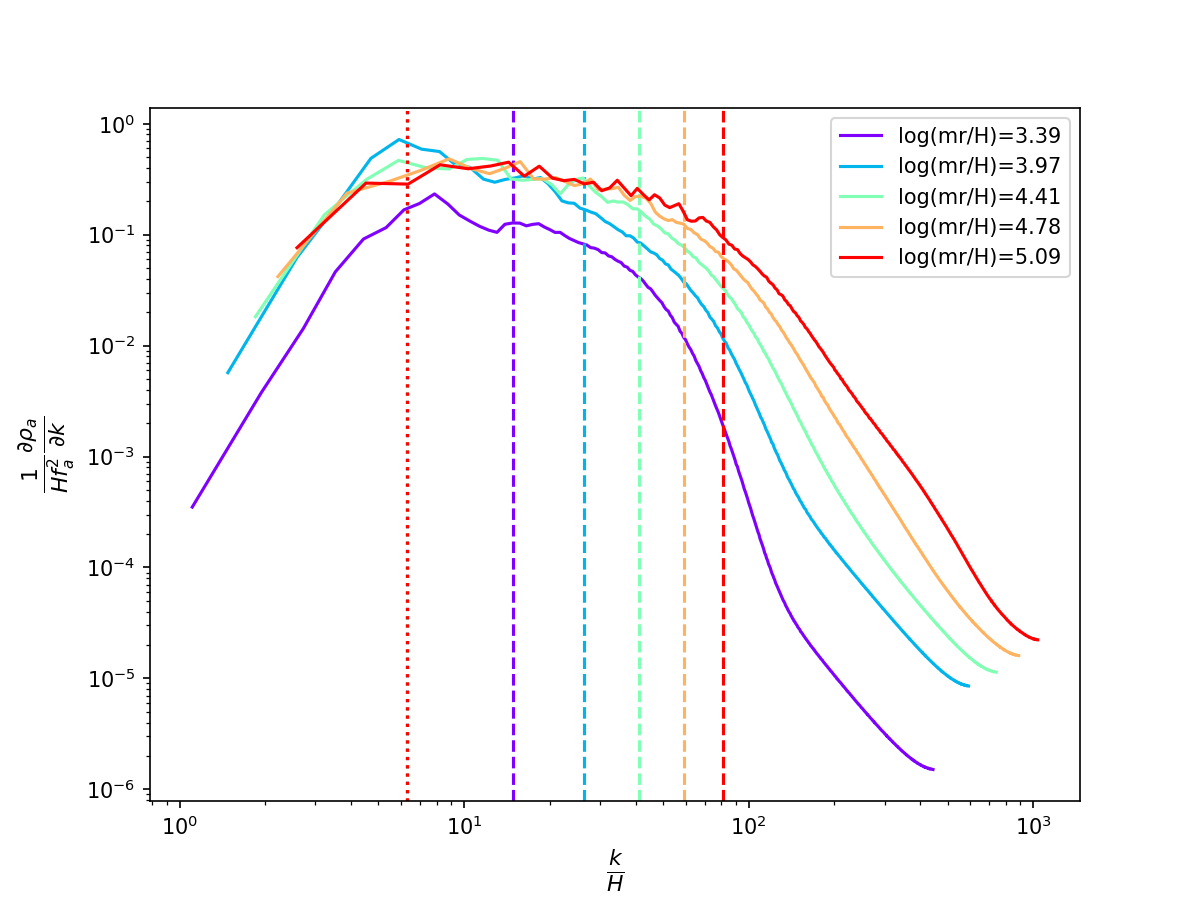}
    \centering
    \includegraphics[width=0.4\textwidth,clip]{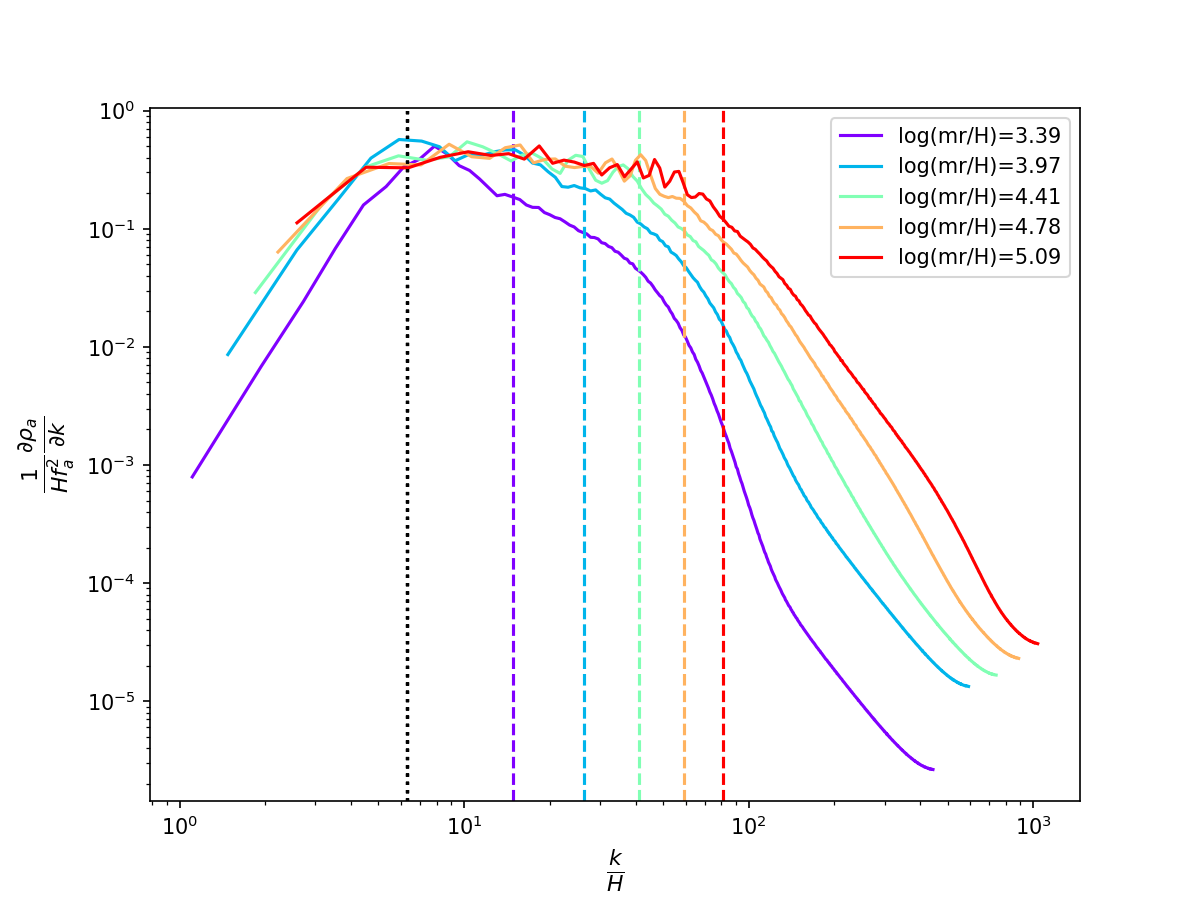} 
    \centering
    \includegraphics[width=0.4\textwidth,clip]{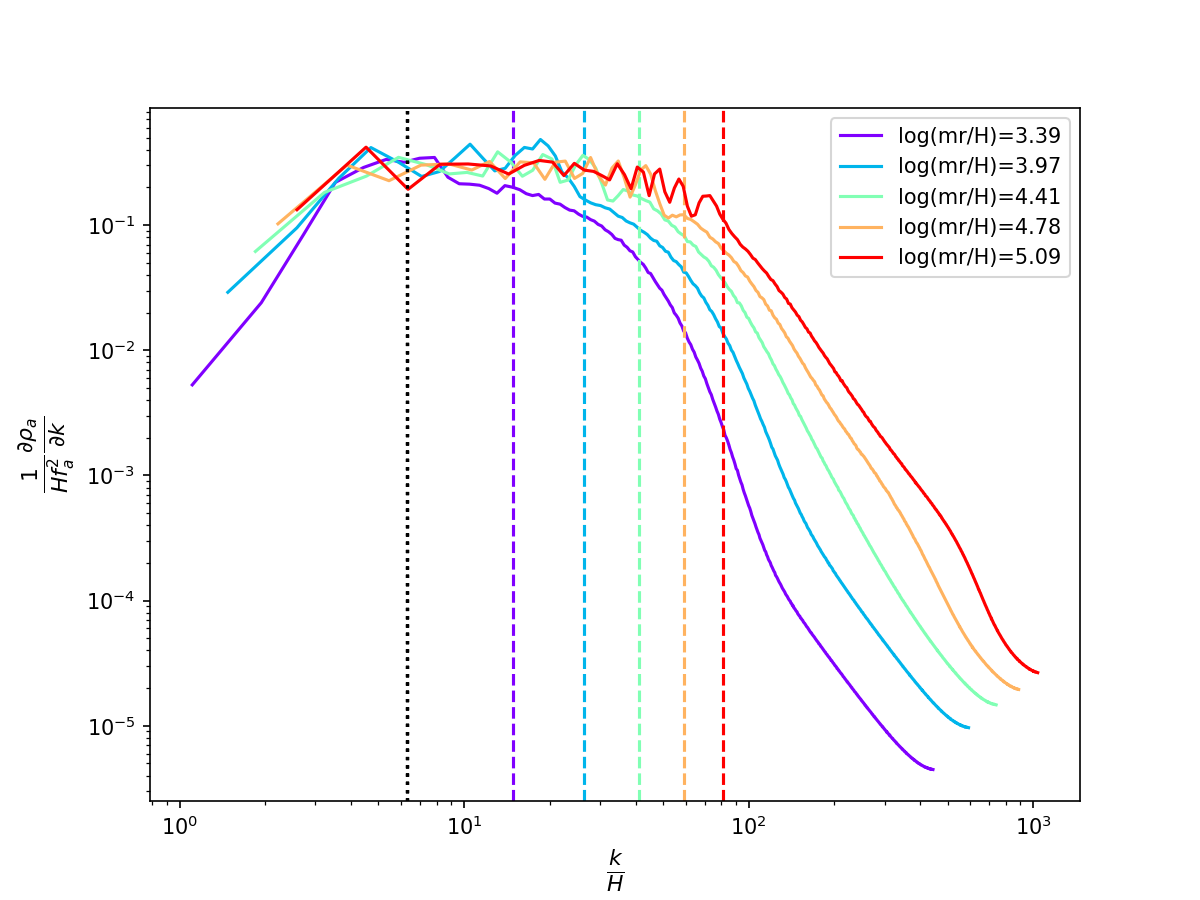}
    \caption{Axion spectrum for Axion-Higgs-1(top), Axion-Higgs-2(middle) and axion(bottom), where the dotted black line corresponds to Hubble length and the dashed line corresponds to $m_r/2$.}
    \label{fig:axionspectrum}
\end{figure}

\begin{figure}[!htp]
    \centering
    \includegraphics[width=0.4\textwidth,clip]{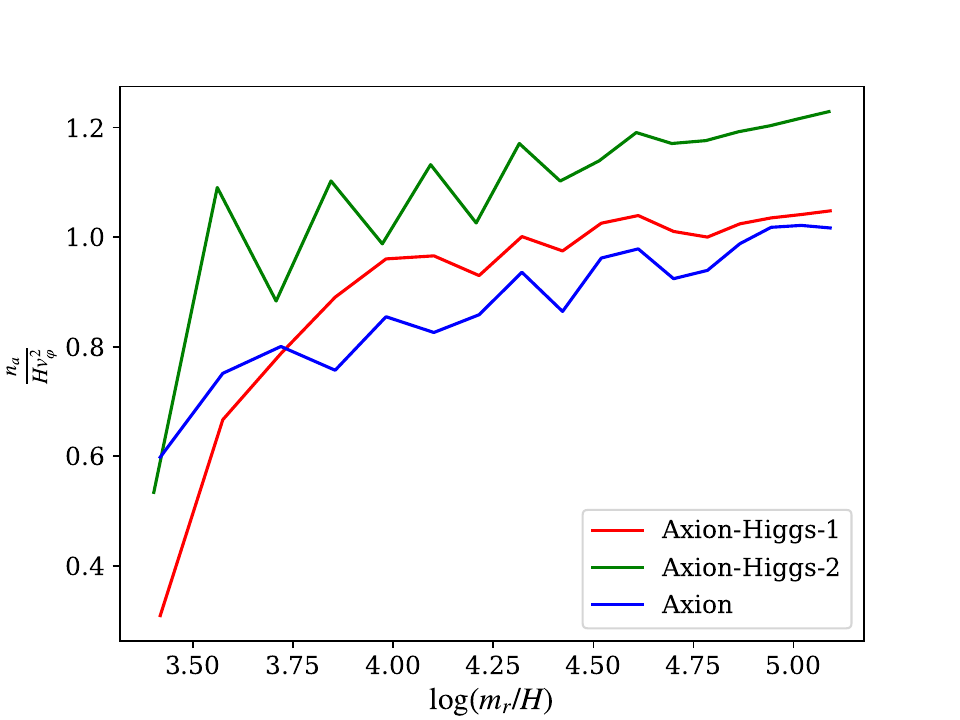}
    \caption{Free axion number density for Axion-Higgs strings and Axion string situations.}\label{fig:n_a}
\end{figure}

The total energy of this system is given by
\begin{align}
    \rho_{total} = \rho_\varphi + \rho_s + \rho_V \;.
\end{align}
, which contains the kinetic and gradient energy of fields and potential energy. By using the form of $\varphi = \frac{r+f_a}{\sqrt{2}}e^{ia/f_a}$, We split the kinetic and gradient energy of complex field into the phase term $\frac{1}{2}\left(\dot{a}^2 + (\nabla a)^2\right)$, radial term $\frac{1}{2}\left(\dot{r}^2+(\nabla r)^2\right)$ and coupled term $(\dot{a}^2 + (\nabla a)^2)(\frac{r}{f_a}+\frac{r^2}{2f_a^2})$. The time derivative of phase is calculated by $\dot{a}=\frac{-\phi_2\dot{\phi_1}+\phi_1\dot{\phi_2}}{\phi_1^2+\phi_2^2}$. \\
To calculate the free axion power spectrum, we should eliminate the contamination from the cosmic string. We use the masked method $\dot{a}_s(x,t) = \dot{a}(x,t)\frac{r+f_a}{f_a}$, where the screening factor $)\frac{r+f_a}{f_a}$ is 0 near string core and 1 far away it. 
The energy density of free axion is given by $\rho_a \approx 2 <\dot{a}^2_{s}> = \int \frac{\partial \rho_a}{\partial |\mathbf{k}|} \mathrm{d}|\mathbf{k}|$, And we get the differential axion spectrum by \cite{Gorghetto_2018}
\begin{equation}
    \frac{\partial \rho_a}{\partial |\mathbf{k}|} = \frac{2|\mathbf{k}|^2}{(2\pi L)^3}\int |\dot{a}_{screen}(\mathbf{k})|^2\mathrm{d}\Omega_\mathbf{k} \;,
\end{equation}
where $\dot{a}_{s}(\mathbf{k})$ is the Fourier transform of  $\dot{a}_{screen}(\mathbf{x})$. 
Fig.\ref{fig:axionspectrum} is shown for the axion spectrum of three cases. We also calculate the number density of free axion by $n_a= \int \frac{1}{ |\mathbf{k}|}\frac{\partial \rho_a}{\partial |\mathbf{k}|} \mathrm{d}|\mathbf{k}|$ for the different case, see Fig.\ref{fig:n_a}.

We divide the relic axion number density calculation into two stages. Firstly, we take the method like \cite{Hiramatsu_2011} to estimate axion number density radiated by string until the time $t_w$, when the temperature-dependent mass of axion is close to Hubble $m_a(t)=3H(t)$ (see definition of axion mass below). Strings energy density is given by $ \rho_s = 2 \pi f_a^2 \ln(t/(\sqrt{\xi(t)}d_s)) \xi(t) / t^2$. The released energy of the string converts into massless axion and the gravitation wave is   
\begin{align}
    \frac{ \mathrm{d}\rho_{str}(t)}{ \mathrm{d}t} = - 2 H(t) \rho_{str}(t) - \left[\frac{\mathrm{d} \rho_{str}(t)}{\mathrm{d}t}\right]_{emi} \;, \\
    \frac{ \mathrm{d}\rho_{a}(t)}{ \mathrm{d}t} = - 4 H(t) \rho_{a}(t) + \left[\frac{\mathrm{d} \rho_{str}(t)}{\mathrm{d}t}\right]_{emi1} \;, \\ \label{eq:enqx}
    \frac{ \mathrm{d}\rho_{gw}(t)}{ \mathrm{d}t} = - 4 H(t) \rho_{gw}(t) + \left[\frac{\mathrm{d} \rho_{str}(t)}{\mathrm{d}t}\right]_{emi2} \;. 
\end{align}
Note that we ignore the equation of gravitation wave by $f_a \ll m_{pl}$. By integrating the differential equations, we get 
\begin{equation}
    N(t)=\int^{t_w}_{t_*} \mathrm{d}t k^{-1}(t) \frac{\mathrm{d}E(t)}{\mathrm{d}t} \;,
\label{eq:axionnumtw}
\end{equation}
where $\bar{N}(t)=a^3n_{a}(t)$ and $\bar{E}(t)=a^4\rho_{a}(t)$ are the number and energy density in comoving volume. The lower limit $t_*$ is the time of string formation defined by $f_a$, where we take the $U(1)$ symmetry breaking temperature $ T \sim f_a $ and $t_*$ is given by $(2t_*)^{-2} = \frac{8 \pi G}{3} \frac{\pi^2}{30} g_* T^4$ with $g_*=106.75$. 

The mean reciprocal comoving momentum of radiated axion $ k^{-1}(t)$ is given by 
\begin{equation}
    k^{-1}(t) = \frac{\int \frac{\mathrm{d}k}{2\pi^2}\frac{1}{k}\Delta P(k,t)}{\int \frac{\mathrm{d}k}{2\pi^2}\Delta P(k,t)}\;,
\end{equation} 
and the differential spectrum is calculated by $\Delta P(k,\eta) = a(t_2)^4P(k,t_2)-a(t_1)^4P(k,t_1)$, where $\rho_a(t)=\int \frac{\mathrm{d}k}{2\pi^2}P(k,t)$ with the comoving momentum $k$. 
Assuming that the physical mean reciprocal momentum is proportional to the Hubble scale, The dimensionless parameters can be extracted by $\epsilon^{-1} = a(t) k^{-1}(t)/(t/2\pi)$ from simulation. We choose $\tilde{\eta_1}=6.0$ corresponding to the string entering the scaling regime and $\tilde{\eta_2}=14.0$ corresponding to the end time of the simulation. The result of $\epsilon^{-1}$ for Axion-Higgs-1, Axion-Higgs-2, and axion strings is 0.87, 0.78, 0.85. As the universe expands around the QCD epoch, the temperature-dependent mass of axion arises, where the mass is given by $m_a=\alpha (T/\Lambda)^{-n} \Lambda^4/f_a^2$ with the parameter $\alpha =1.68*10^{-7}, \Lambda = 400 \text{MeV}, n = 6.68$ from Ref.~\cite{Wantz_2010}. 
Finally, without considering the influence of axion number density caused by domain wall, We directly redshift $n(t_w)$ to today, where the scale factor ratio $(a_0/a_w)^{-3}$ is given by Eq.\ref{eq:scalefactor} with $g_{*w}=70$. 
In addition, the axion relic density by misalignment mechanism is given by $\Omega_{a}=4.63*10^{-3}*(f_a/10^{10}{\rm GeV})^{(6+n)/(4+n)}(\Lambda/400{\rm MeV})$~\cite{PhysRevD.91.065014}. We get the conservative upper bound for the symmetry of $f_a < 5.1*10^9, 5.6*10^9, 3.7*10^9 {\rm GeV}$ for Axion-Higgs-1, Axion-Higgs-2 and axion string cases, with the limitation being $\Omega_{DM}h^2<0.12$.

\section{Gravitational wave power spectrum}
\label{app:gws power spectrum}
\begin{figure}[!htp]
    \centering
    \includegraphics[width=0.4\textwidth,clip]{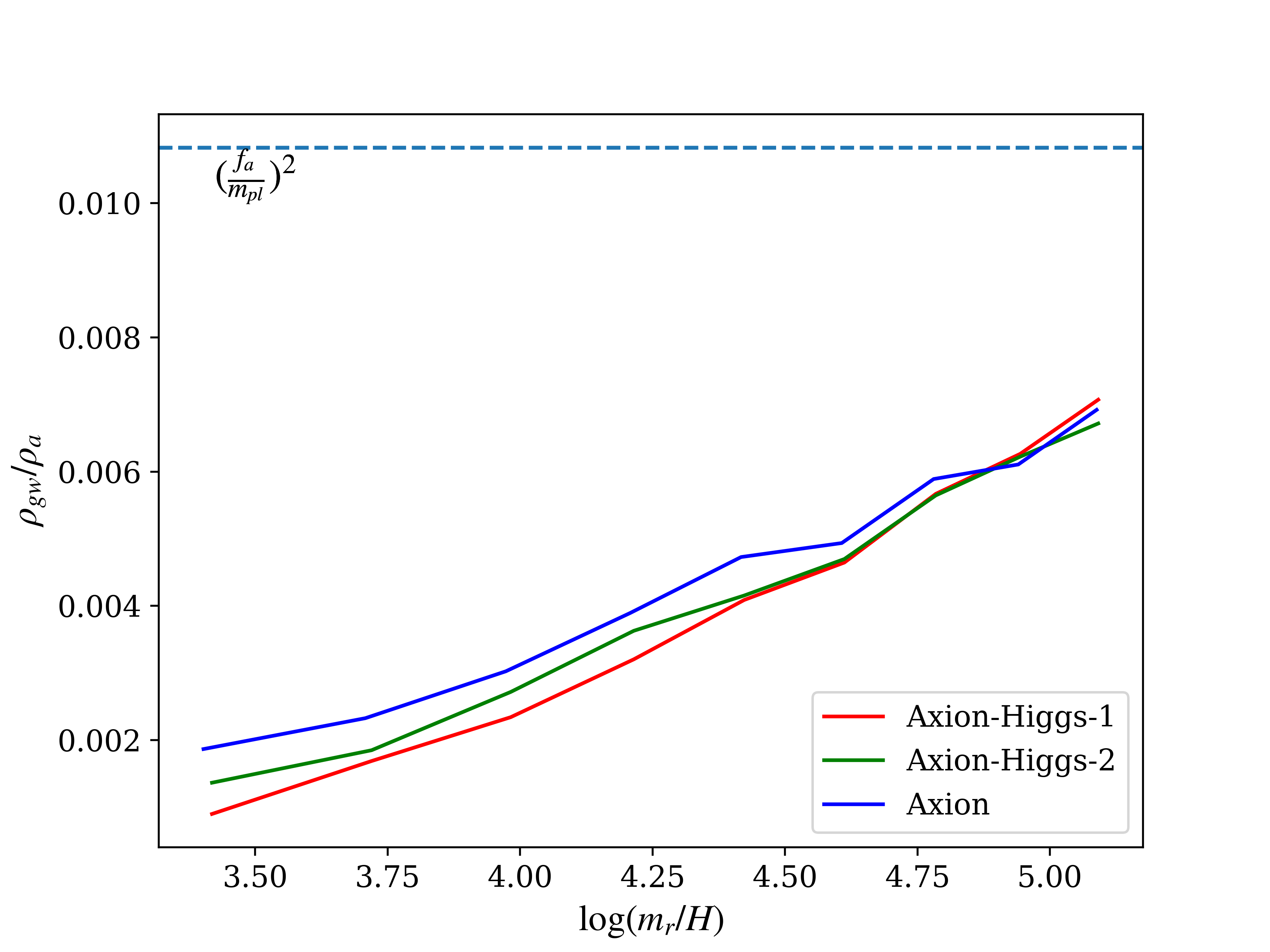}
    \caption{The ratio of $\rho_{gw}$ to $\rho_{a}$ for three cases.}
    \label{fig:rhoarhoagw}
\end{figure}

GWs are the transverse and traceless(TT) part of the metric perturbation $h_{ij}$ with $ h_{ij} = g_{ij} - \delta_{ij}$. 
The equation of motion of GWs is
\begin{equation}
    \ddot{h}_{ij} + 3H\dot{h}_{ij} - \frac{1}{a^2} \nabla^2 h_{ij} = \frac{2}{m_{pl}^2a^2}\Pi^{TT}_{ij}	
    \label{eq:gws} \;,
\end{equation}
where $\dot{\quad}$ is the derivative of cosmic time, $m_{pl}$ is the reduced Planck mass, and $\Pi^{TT}_{ij}$ is the transverse and traceless part of anisotropic tensor $\Pi_{ij}$. 
The anisotropic tensor is given by $\Pi_{ij} = T_{ij} - pg_{ij}$, where the energy momentum tensor is calculated by $T_{ij} = \partial_i s \partial_j s + 2 \sum_{k=0,1} \partial_i \varphi_k \partial_j \varphi_k$, and the second term background pressure does not contribute to $\Pi^{TT}_{ij}$.  
The transverse and traceless anisotropic tensor is obtained by $\Pi_{ij}^{TT} = \Lambda_{ij,kl} \Pi_{kl}$ in Fourier space, and the projection operator is defined as
\begin{align}
    \Lambda_{ij,kl}(\hat{\textbf{k}}) &= P_{ik}({\hat{\textbf{k}}}) P_{jl}({\hat{\textbf{k}}})-\frac{1}{2}P_{ij}({\hat{\textbf{k}}})P_{kl}({\hat{\textbf{k}}}) \;,\nonumber \\
    P_{ij}({\hat{\textbf{k}}}) &= \delta_{ij} - \hat{k}_i\hat{k}_j \;,
\end{align}
where $\hat{k}_i = k_i / k$ is the unit vector in $\textbf{k}$ direction. 
For the convenience of numerically calculating GWs in lattice simulation, the auxiliary tensor $u_{ij}$ is introduced and evolved in fact, where the metric perturbation can be calculated by $h_{ij} = \Lambda_{ij,kl} u_{ij}$. The equation of motion of $u_{ij}$ is given by
\begin{align}
    \ddot{u}_{ij} + 3H\dot{u}_{ij} - \frac{1}{a^2} \nabla^2 u_{ij} = \frac{2}{m_{pl}^2a^2}\Pi_{ij} \;.
\end{align}
The gravitational wave energy density is defined as 
\begin{align}
    \rho_{GW}(t) = \frac{1}{32 \pi G}\left<\hat{h}_{ij}(\textbf{x},t) \hat{h}_{ij}(\textbf{x},t) \right> \;, 
\end{align}
and $\left<...\right>$ is the spatial average. We show the value of $\rho_{gw}/ \rho_a$ in Fig.\ref{fig:rhoarhoagw}, which is slightly smaller than the same order of magnitude as $(f_a/m_{pl})^2$ during the simulation regime. Ref.~\cite{Baeza-Ballesteros:2023say} found the ratio between the emission rates of GWs and particles is of the order $\mathcal{O}(10)(f_a/m_{pl})^2$.

\begin{figure}[!htp]
    \centering
    \includegraphics[width=0.4\textwidth,clip]{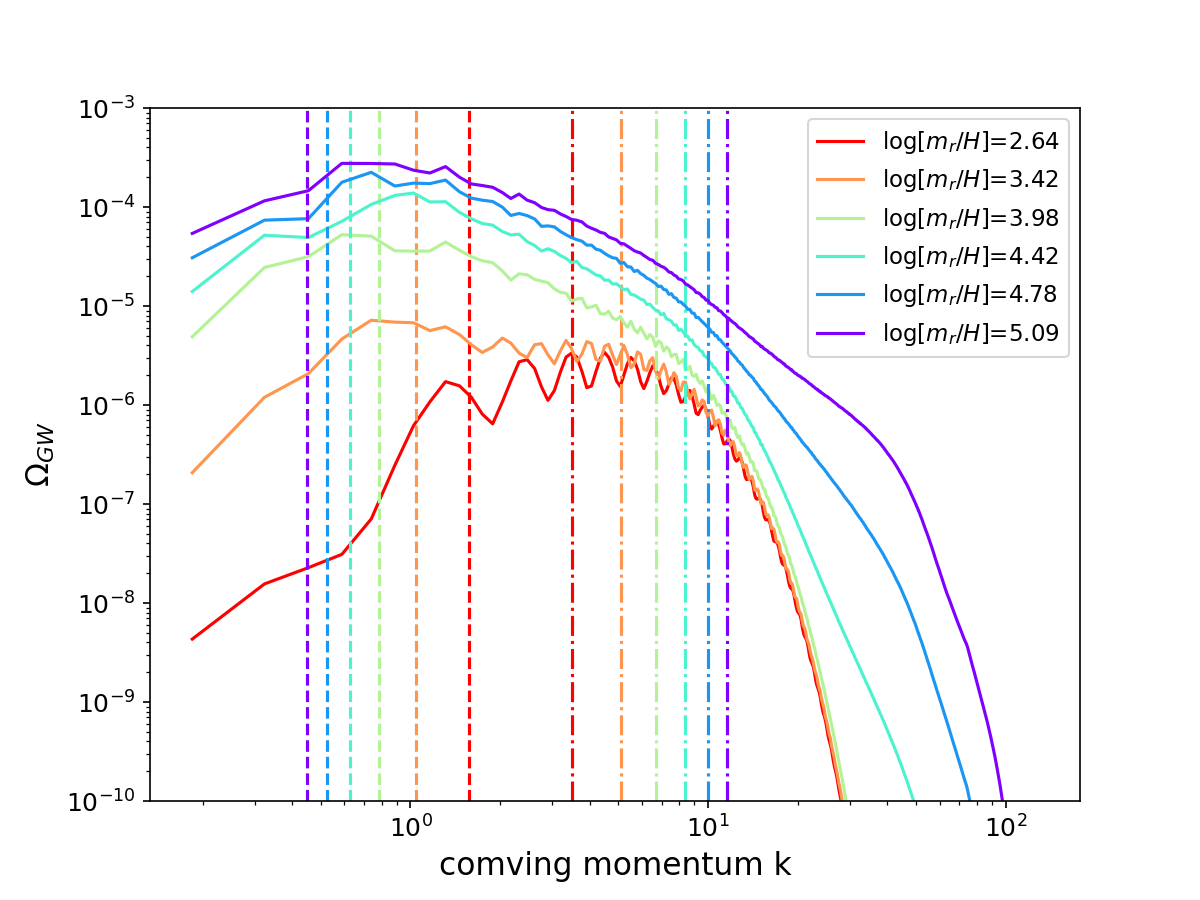}
    \centering
    \includegraphics[width=0.4\textwidth,clip]{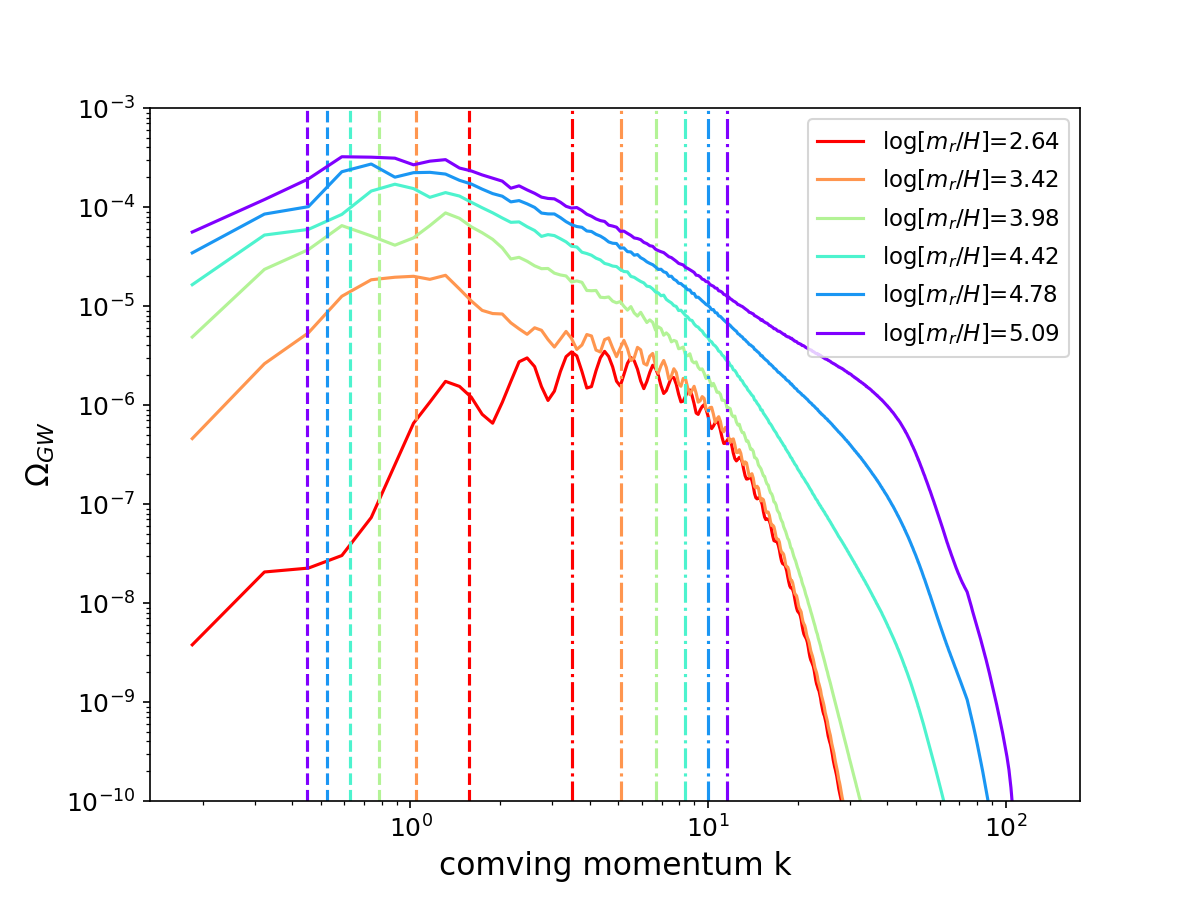} 
    \centering
    \includegraphics[width=0.4\textwidth,clip]{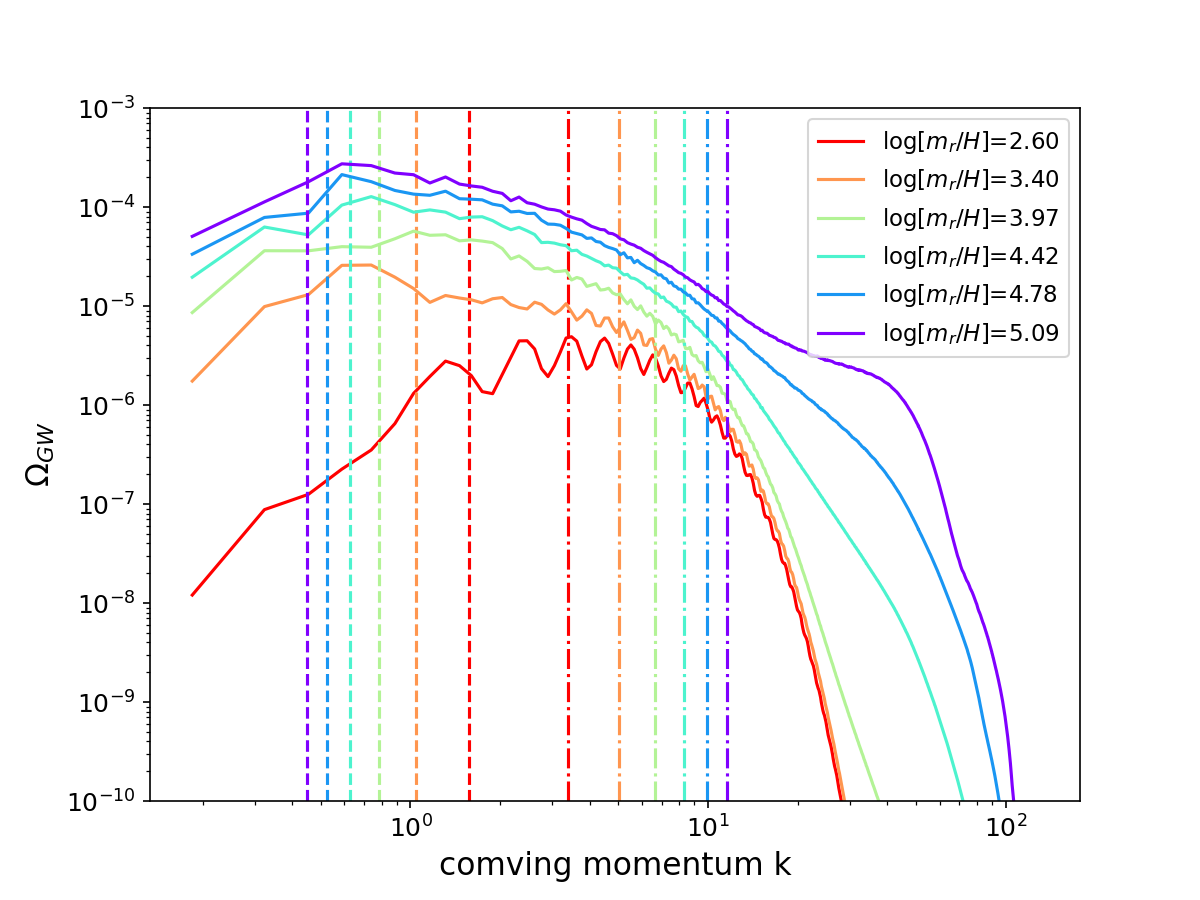}
    \caption{GWs spectra for global strings of Axion-Higgs-1 (top), Axion-Higgs-2 (middle) and Axion  (bottom), where the left dashed line corresponds to Hubble length $2\pi H$ and the dashed-dotted line corresponds to $m_r$.}
    \label{fig:allgwsspectra}
\end{figure}

The gravitation wave power spectrum in momentum space is given by 
\begin{align}
    \frac{\mathrm{d} \rho_{GW}}{\mathrm{d} \log{k}} = \frac{k^3}{(4\pi)^3G} P_{\hat{h}}(k,t) \;,
\end{align}
where $\left<\hat{h}_{ij}(\textbf{k},t) \hat{h}_{ij}(\textbf{k},t) \right> = (2\pi)^3P_{\hat{h}}(k,t) \delta^{(3)}(\textbf{k} - \textbf{k}^\prime)$. 
We present the gravitational wave power spectrum in Fig. \ref{fig:allgwsspectra}, where $\Omega_{GW}$ is the GWs power spectrum normalized by the critical density $\rho_c = \frac{3H^2}{8\pi G}$. The IR peak around the Hubble scale arises during the formation of string and enhances persistently during the scaling regime, which corresponds to the typical string inter-distance.  
We take the GWs power spectrum redshift method used in Ref.\cite{Price_2008}. The GWs energy density is scaled as $ a^{-4}$ like radiation. Assuming the entropy is conserved, the redshift factor is calculated by 
\begin{equation}
    \frac{a_0^4}{a_e^4}=\frac{g_e^{1/3}\rho_{rad,e}}{g_0^{1/3}\rho_{rad,0}} \;.
    \label{eq:scalefactor}
\end{equation}
where the subscript 0 and e denote today and the time of GWs should be reshifted. Here We take $g_0=3.36$ and $\rho_{rad,e}=\rho_c$ in the background of the radiation-dominant universe.

\bibliography{reference}
\end{document}